\documentclass[prc,aps,eqsecnum,floatfix,showpacs]{revtex4}
\usepackage{dcolumn}
\usepackage{bm}
\usepackage{graphicx}
\begin{document}

\title{Two-Body Electrodisintegration of $^3$He at High Momentum Transfer} 
\author{R.\ Schiavilla}
\affiliation{Jefferson Lab, Newport News, Virginia 23606 \\
         Department of Physics, Old Dominion University, Norfolk, Virginia 23529, USA}
\author{O.\ Benhar}
\affiliation{Istituto Nazionale di Fisica Nucleare and Dipartimento di Fisica,
Universit\`a \lq\lq La Sapienza\rq\rq, I-00185 Roma, Italy}
\author{A.\ Kievsky, L.E.\ Marcucci, and M.\ Viviani}
\affiliation{Istituto Nazionale di Fisica Nucleare and Dipartimento di Fisica,
Universit\`a di Pisa, I-56100 Pisa, Italy}
\date{\today}

\begin{abstract}
The $^3$He$(e,e^\prime p)$$d$ reaction is studied using an
accurate three-nucleon bound state wave function, a model
for the electromagnetic current operator including one-
and two-body terms, and the Glauber approximation for the
treatment of final state interactions.  In contrast to earlier studies,
the profile operator in the Glauber expansion is derived from a
nucleon-nucleon scattering amplitude, which retains its full
spin and isospin dependence and is consistent with phase-shift
analyses of two-nucleon scattering data.  The amplitude is
boosted from the center-of-mass frame, where parameterizations
for it are available, to the frame where rescattering occurs.
Exact Monte Carlo methods are used to evaluate the relevant
matrix elements of the electromagnetic current operator.  The
predicted cross section is found to be in quantitative agreement
with the experimental data for values of the missing momentum
$p_{\rm m}$ in the range (0--700) MeV/c, but underestimates the
data at $p_{\rm m} \simeq 1$ GeV/c by about a factor of two.
However, the longitudinal-transverse asymmetry, measured up
to $p_{\rm m} \simeq$ 600 MeV/c, is well reproduced by theory.
A critical comparison is carried out between the results obtained
in the present work and those of earlier studies.
\end{abstract}
\maketitle
\section{Introduction}
\label{sec:intro}

Recent experiments at JLab have yielded beautiful data for the
$^3$He$(e,e^\prime p)$$d$ cross section and longitudinal-transverse
asymmetry up to missing momenta $p_{\rm m} \simeq 1.1$ and 0.6 GeV/c~\cite{Rvachev05},
respectively, and the three-body breakup cross
section $^3$He$(e,e^\prime p)$$p$$n$ for missing momenta and energies
($E_{\rm m}$) in the range $p_{\rm m}$ up to 0.84 GeV/c and $E_{\rm m}$
up to 140 MeV~\cite{Benmokhtar05}.  These data have spurred renewed interest
in these reactions, which has led to a series of
papers~\cite{Ciofi05,Ciofi05a,Laget05,Laget05a}, dealing with the theoretical
description of the proton-knockout mechanism and, in particular, with
the treatment of final state interactions (FSI) at GeV energies. 

In the present work, we contribute to this effort.  As in
Refs.~\cite{Ciofi05,Ciofi05a} we adopt the Glauber approximation to describe
the rescattering processes between the struck proton and the nucleons in
the recoiling deuteron.  However, in contrast to the studies of
Refs.~\cite{Ciofi05,Ciofi05a,Laget05,Laget05a},
we retain the full spin and isospin dependence
of the nucleon-nucleon ($N$$N$) scattering
amplitude from which the Glauber profile operator
is derived, and do not make use of the factorization approximation, which
allows one to write the $(e,e^\prime p)$ cross section on a nucleus
in terms of a proton cross section times a distorted spectral function.
A number of issues pertaining to the parameterization of the $N$$N$ amplitude,
and its boosting from the center-of-mass (c.m.) frame to the frame in which
rescattering occurs, are discussed in considerable detail.

The spin dependence of the $N$$N$ amplitudes will turn out to
play an important role in the high $p_{\rm m}$ region where double
rescattering effects become dominant.  It is not clear that one is
justified in ignoring it, particularly in the analysis of experiments,
such as the $^4$He$(\vec e,e^\prime \vec p\, )$$^3$H reaction~\cite{Strauch03},
in which the polarizations of the ejected proton are measured~\cite{Lava04}.

The present paper is organized as follows.  In Sec.~\ref{sec:wcss} we
briefly review the bound-state wave functions and the model for the
electromagnetic current operator, while in Sec.~\ref{sec:glb} the Glauber
approximation, the profile operator, and the parameterization for
the $N$$N$ scattering amplitude are discussed.  Next, in Sec.~\ref{sec:form},
the relevant formulae for the $(e,e^\prime p)$ cross section and their
limits in the plane-wave-impulse-approximation (PWIA) are summarized,
while in Sec.~\ref{sec:calc} the Monte Carlo method, as implemented
in the present calculations, is described.  Finally, in Sec.~\ref{sec:res}
we present a detailed discussion of the results, including a critical
comparison between the present and earlier studies, while
in Sec.~\ref{sec:concl} we summarize our conclusions.
\section{Wave functions and currents}
\label{sec:wcss}

In this section we briefly describe the $^3$He wave function
and the model for the nuclear electromagnetic current.
The discussion is rather cursory, since both these aspects of the
calculations presented in this work have already been reviewed in
considerable detail in a number of earlier papers.  References
to these are included below.
\subsection{Bound-state wave functions}
\label{sec:wf}

The ground states of the $A$=3 nuclei are represented by variational
wave functions, derived from a realistic Hamiltonian consisting of the
Argonne $v_{18}$ two-nucleon ($N$$N$)~\cite{Wiringa95} and Urbana-IX
three-nucleon ($N$$N$$N$)~\cite{Pudliner95} interactions---the AV18/UIX
Hamiltonian model---with the correlated hyperspherical-harmonics
(CHH) method~\cite{Kievsky93}.  The high accuracy of the CHH wave
functions is well documented~\cite{Nogga03}, as is the quality of the
AV18/UIX Hamiltonian in successfully and quantitatively accounting for
a wide variety of three-nucleon bound-state properties and reactions,
ranging from binding energies, charge radii, and elastic form
factors~\cite{Nogga03,Marcucci98,Carlson98} to low-energy radiative and weak
capture cross sections and polarization observables~\cite{Marcucci04},
to the quasi-elastic response in inclusive $(e,e^\prime)$ scattering at
intermediate energies~\cite{Carlson00}.
\subsection{Electromagnetic current operator}
\label{sec:cnt}

The nuclear electromagnetic current includes one- and two-body terms,

\begin{eqnarray}
\rho({\bf q})&=&\sum_{i=1}^A \rho_i({\bf q}) + \sum_{i<j=1}^A \rho_{ij}({\bf q}) \ , \\
{\bf j}({\bf q})&=&\sum_{i=1}^A {\bf j}_i({\bf q}) + \sum_{i<j=1}^A {\bf j}_{ij}({\bf q}) \ .
\end{eqnarray}
The one-body current and charge operators have the form recently derived
by Jeschonnek and Donnelly~\cite{Jeschonnek98} from an expansion of the
covariant single-nucleon current, in which only terms dependent
quadratically on the initial nucleon momentum (and higher order terms)
are neglected.  In momentum space, they are explicitly given by

\begin{equation}
\rho_i ({\bf q})=\frac{q}{Q} G_E + \frac{\rm i}{\sqrt{1+\eta}} \frac{1}{2m^2}
\left( G_M-\frac{1}{2} G_E\right) {\bm \sigma}_i  \cdot ({\bf q} \times {\bf p}_i ) \ , 
\label{eq:rho1}
\end{equation}
\begin{equation}
{\bf j}_i  ({\bf q})=\frac{Q}{q} \left[ 
\frac{ {\bf p}_i} {m} \left( G_E +\frac{\eta}{2} G_M \right)
-\frac{\rm i}{2m}  G_M \left( {\bf q} \times {\bm \sigma}_i + \frac{\omega}{2m}
{\hat {\bf q}} \cdot {\bm \sigma}_i \, \hat{\bf q} \times {\bf p}_i \right) \right] \ ,
\end{equation}
where ${\bf q}$ ($\omega$) is the virtual photon three-momentum (energy)
transfer, $Q$ is the four-momentum transfer with $Q^2$=$q^2-\omega^2$
$>0$, $\eta$ is defined as $\eta\equiv Q^2/(4m^2)$, $m$ being the nucleon
mass, and ${\bf p}_i$ and ${\bm \sigma}_i$ are the momentum
and spin operators of nucleon $i$, respectively.  The nucleon Sachs form factors
$G_E$ and $G_M$ are defined as 

\begin{equation}
G_{E/M} = \frac{1}{2} \left[              ( G_{Ep/Mp} + G_{En/Mn} ) 
                            +\tau_{i,z} \,( G_{Ep/Mp} - G_{En/Mn} ) \right] \ ,
\end{equation}
where the $Q^2$ dependence is understood, and $\tau_z$ is the
$z$-component of the isospin.  The H\"ohler parameterization~\cite{Hohler76}
is used for the proton and neutron electric and magnetic form factors
$G_{Ep/En}$ and $G_{Mp/Mn}$.

The form adopted above for the one-body currents is well suited for dealing
with processes in which the energy transfer may be large (i.e., the ratio
of four- to three-momentum transfer $(Q/q)^2$ is not close to one)
and the initial momentum of the nucleon absorbing the virtual photon
is small.  Thus, its use is certainly justified in quasi-elastic kinematics
for moderate values of the missing momentum.  Note that in the limit
$(Q/q)^2 \simeq 1$ one recovers the standard non-relativistic expressions for the
impulse-approximation currents (including the spin-orbit correction
to the charge operator).
 
The two-body charge, $\rho_{ij}({\bf q})$, and current, ${\bf j}_{ij}({\bf q})$,
operators consist of a \lq\lq model-independent\rq\rq part, that is constructed
from the $N$$N$ interaction (the AV18 in the present case), and a
\lq\lq model-dependent\rq\rq one, associated with the excitation of nucleons to
$\Delta$ resonances (for ${\bf j}_{ij}$ only)
and $\rho\pi\gamma$ and $\omega\pi\gamma$ transitions
(for a review, see Ref.~\cite{Carlson98} and references therein).
Improvements in the construction of the model-independent two-body currents originating
from the momentum-dependent terms of the $N$$N$ interaction have been recently
reported in Ref.~\cite{Marcucci05}.  In this latter work, three-body currents
associated with $N$$N$$N$ interactions have also been derived.  Both these
refinements, however, are expected to have little impact on the results of
the present study, and therefore are not considered any further.

The present model for two-body charge and current
operators is quite realistic at small momentum transfers.
However, for processes involving momentum and energy
transfers of order 1 GeV, such as the $^3$He($e,e^\prime p$)$d$
reaction under consideration here for which $q$=1.50 GeV/c and $\omega$=0.84 GeV,
it is likely to have additional corrections.
\section{Final state interactions: Glauber approximation}
\label{sec:glb}

In the kinematics considered in Sec.~\ref{sec:res}, the proton lab kinetic
energies are typically of the order 0.5 GeV or larger.  These energies
are obviously beyond the range of applicability of $N$$N$ interaction
models, such as the AV18, which are constrained to reproduce $N$$N$
elastic scattering data up to the pion production threshold.  At higher energies,
$N$$N$ scattering becomes strongly absorptive with the opening of particle
production channels.  Indeed, the $p$$p$ inelastic cross section at 0.5 GeV
increases abruptly from about 2 mb to 30 mb, and remains essentially constant
for energies up to several hundred GeV~\cite{Lechanoine93}.
 
On the other hand, the small momentum transfer which characterizes
scattering processes at high energies makes the Glauber approximation~\cite{Glauber59}
particularly well suited in this regime.  Another advantage is its reliance
on a $N$$N$ scattering amplitude, which is fitted to data.  It is the approach
we adopt in the present work to describe the wave function of the final $p$+($A-1$)
system as 

\begin{equation}
\psi(p+^{(A-1)}\!f;{\rm GLB})=
\frac{1}{\sqrt{A}} \sum_{\cal P} \epsilon_{\cal P}\, G(A;1 \dots A-1)\,
{\rm e}^{{\rm i} {\bf p} \cdot {\bf r}_A} \chi_\sigma(A;p)\,
       {\rm e}^{{\rm i} {\bf p}_f \cdot {\bf R}_{1\dots A-1}}
\phi_{\sigma_f}(1\dots A-1;f) \ ,
\label{eq:glb}
\end{equation}
where $\chi_\sigma(p)$ represents a proton in spin state $\sigma$,
$\phi_{\sigma_f}(f)$ denotes the wave function of the ($A-1$)-system
with spin projection $\sigma_f$, and ${\bf R}_{1 \dots A-1}$ is the
center-of-mass position vector of the $A-1$ nucleons in this cluster.
The sum over permutations $\cal P$ of parity $\epsilon_{\cal P}$ ensures
the overall antisymmetry of $\psi(p+^{(A-1)}\!f;{\rm GLB})$.

The operator $G(A;1 \dots A-1)$ inducing final-state-interactions (FSI)
can be derived from an analysis of the multiple scattering series by
requiring that the struck (fast) nucleon (nucleon $A$) moves in a
straight-line trajectory (that is, it is undeflected by rescattering
processes), and that the nucleons in the residual system (nucleons
$1, \dots, A-1$) act as fixed scattering centers~\cite{Wallace81,Benhar00}
(the so-called {\it frozen approximation}).  It is expanded as

\begin{equation}
G=1 + \sum_{n=1}^{A-1} (-)^n G^{(n)} \ ,
\end{equation}
where $G^{(n)}$ represents the $n^{\rm th}$ rescattering term, and therefore
for an $A$-body system up to $A-1$ rescattering terms are generally present.
The leading single-rescattering term reads

\begin{equation}
G^{(1)}(A;1\dots A-1)=\sum_{i=1}^{A-1} \theta(z_{iA})
\, \Gamma_{iA}({\bf b}_{iA};s_{iA}) \ ,
\label{eq:glb1}
\end{equation}
where $z_{iA}$ and ${\bf b}_{iA}$ denote the longitudinal and
transverse components of ${\bf r}_i-{\bf r}_A$ relative to
$\hat {\bf p}$, the direction of the nucleon momentum,

\begin{equation}
z_{iA} \equiv {\hat{\bf p}}\cdot ({\bf r}_i-{\bf r}_A) \ , \qquad
{\bf r}_i-{\bf r}_A \equiv {\bf b}_{iA} + z_{iA} \,{\hat{\bf p}} \ ,
\end{equation}
and the step-function $\theta(x)$, $\theta(x)=1$ if $x>0$, prevents
the occurrence of backward scattering for the struck nucleon.  The
\lq\lq profile operator\rq\rq $\, \Gamma_{iA}$, to be discussed below,
is related to the Fourier transform of the $N$$N$ scattering amplitude
at the invariant energy $\sqrt{s_{iA}}$.

The double-rescattering term, relevant for the present study of the
$^3$He($e,e^\prime p$)$d$ reaction, is given by

\begin{equation}
G^{(2)}(A;1\dots A-1)=\sum_{i\ne j=1}^{A-1} 
\theta(z_{ij})\, \theta(z_{jA}) \,
  \Gamma_{iA}({\bf b}_{iA};s_{iA})\, \Gamma_{jA}({\bf b}_{jA};s_{jA}) \ ,
\end{equation}
where the product of $\theta$-functions ensures the correct sequence 
of rescattering processes in the forward hemisphere.  For example, if
$z_j-z_A > 0$ and $z_i-z_j >0$, then nucleon $A$ scatters first from
nucleon $j$ and then from nucleon $i$.  Note that the operators
$\Gamma_{iA}$ and $\Gamma_{jA}$ do not generally commute.
\subsection{The profile operator $\Gamma_{ij}$}
\label{sec:gamma}

\begin{figure}[bthp]
\includegraphics[width=6in]{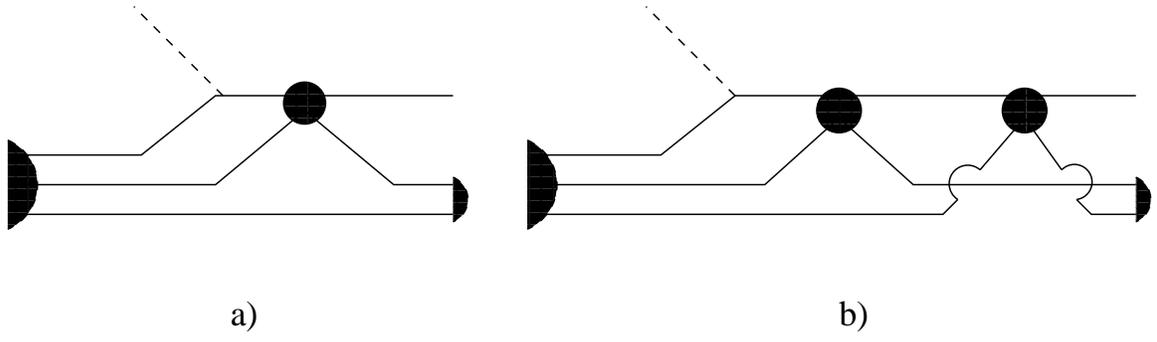}
\caption{Schematic illustration of single, panel a), and double, panel b),
rescattering processes.  Dashed (solid) lines represent virtual photons (nucleons),
while black solid circles represent $N$$N$ scattering amplitudes.}
\label{fig:glb}
\end{figure}

At this stage it is useful to specify the kinematics of the
various rescattering processes occurring in the Glauber expansion.
Specializing to the $^3$He($e,e^\prime p$)$d$ reaction of interest
here, the single- and double-rescattering terms are illustrated schematically
in Fig.~\ref{fig:glb}.  In this figure, nucleon 3 denotes the knocked-out
nucleon with momentum ${\bf p}_3$=${\bf p}$ and energy
$E_3$=$E$ in the lab frame, while nucleons 1 and 2, making up the
deuteron, have momenta ${\bf p}_1$ and ${\bf p}_2$,
respectively, with ${\bf p}_1$+${\bf p}_2$=${\bf p}_d$ (again,
in the lab frame).  The black solid circle represents the $N$$N$ scattering
amplitude.  In the single-rescattering case, the $N$$N$ amplitudes in
the two terms (panel a) of Fig.~\ref{fig:glb} only shows one of them)
are evaluated at the invariant energies $\sqrt{s_{i3}}$, $i$=1,2, with

\begin{eqnarray}
s_{i3} &=& (E_i + E_3)^2-({\bf p}_i +{\bf p}_3)^2  \nonumber \\
      &\simeq & 2\, m^2+E \, \sqrt{ {\bf p}_d^2+ 4\, m^2 }- {\bf p} \cdot {\bf p}_d \ ,
\label{eq:kin}
\end{eqnarray} 
where in the second line it has been assumed that i) the nucleons
are on their mass shells, and ii) nucleons 1 and 2 in the recoiling
deuteron share its momentum equally, ${\bf p}_i \simeq {\bf p}_d/2$.
The momenta of nucleon 3 and nucleon $i$, $i$=1,2, after rescattering
are ${\bf p}-{\bf k}$ and ${\bf p}_d/2+{\bf k}$, where ${\bf k}$
denotes the momentum transfer.  The spectator nucleon ($j \not= i$)
has momentum ${\bf p}_d/2$.  Thus, the \lq\lq rescattering frame\rq\rq
we refer to in the following is defined as that in which nucleon 3
(nucleon $i$) have initial and final momenta ${\bf p}$ and
${\bf p}-{\bf k}$ (${\bf p}_d/2$ and ${\bf p}_d/2+{\bf k}$), respectively.

In the case of double rescattering, panel b) of Fig.~\ref{fig:glb},
a similar analysis can be carried out.  In particular, it leads in the
eikonal limit to the approximation in which both $N$$N$ amplitudes are
evaluated at the invariant energies $\sqrt{s_{13}} \simeq \sqrt{s_{23}}$,
as obtained in Eq.~(\ref{eq:kin}).
 
The profile operator $\Gamma_{ij}$ is related to the $N$$N$ scattering amplitude
in the rescattering frame, denoted as $F_{ij}({\bf k};s)$, via the Fourier transform 

\begin{equation}
\Gamma_{ij}({\bf b};s) = \frac{1}{ 2 \pi {\rm i}\, p }
\int {\rm d}^2{\bf k}\, {\rm e}^{-{\rm i} {\bf k}\cdot {\bf b}} F_{ij}({\bf k};s) \ ,
\label{eq:prof}
\end{equation} 
where, in the eikonal limit, the momentum transfer ${\bf k}$ is
perpendicular to ${\bf p}$.  The isospin symmetry of the strong interactions
allows one to express $F_{ij}$ as 

\begin{equation}
F_{ij}=F_{ij, +}+F_{ij,-} {\bm \tau}_i \cdot {\bm \tau}_j \ ,
\end{equation}
where the $F_{ij,\pm}$ are related to the physical amplitudes
for $p$$p$ and $p$$n$ scattering via

\begin{equation}
F_{ij,\pm}=\frac{ F^{pp}_{ij} \pm F^{pn}_{ij} }{2} \ .
\end{equation}
Available parameterizations of the $p$$p$ and $p$$n$ amplitudes are
given in the c.m. frame~\cite{Wallace81}, and therefore
one needs to boost these from the c.m.~to the rescattering frame.  In the discussion
that follows, the general form of the c.m.~amplitude and its parameterization
are described first, while the boosting procedure
adopted in the present work is illustrated next.
\subsection{The $N$$N$ scattering amplitude in the c.m.~frame}
\label{sec:fcm}

It is well known that the most general form for the $N$$N$ scattering amplitude
in the c.m.~frame reads (see Refs.~\cite{Lechanoine93,Wallace81} and references
therein)

\begin{equation}
(2 {\rm i}\, \overline{p})^{-1}\, \overline{F}^{\, NN}_{ij}(\,\overline{{\bf k}},s)
=\sum_{m=1}^5 \overline{F}^{\, NN}_m(s,\overline{{\bf k}}^{\, 2}) \overline{O}^{\, m}_{ij} \ ,
\label{eq:fnn}
\end{equation}
where $\overline{\bf p}$ and $\overline{\bf p}^{\, \prime}$ denote the initial
and final nucleon momenta, respectively, the $\overline{F}^{\, NN}_m$'s are functions
of the invariant energy $\sqrt{s}$ and momentum transfer $\overline{\bf k}^{\, 2}$
(with $\overline{\bf k}=\overline{\bf p}-\overline{\bf p}^{\, \prime}$), and a
possible choice for the five operators $\overline{O}^{\, m}_{ij}$, especially
convenient for our present purposes, is that given in Ref.~\cite{Wallace81},

\begin{equation}
\overline{O}^{\,m=1,\dots,5}_{ij}=1\, ,\,  {\bm \sigma}_i \cdot {\bm \sigma}_j\, ,\, 
{\rm i}\, ({\bm \sigma}_i+{\bm \sigma}_j) \cdot \overline{\bf k} \times {\overline{\bf e}}\, ,\,
{\bm \sigma}_i \cdot \overline{\bf k} \, {\bm \sigma}_j \cdot \overline{\bf k}\, ,\,
{\bm \sigma}_i \cdot {\overline{\bf e}} \, {\bm \sigma}_j \cdot {\overline{\bf e}} \ .
\label{eq:opnn}
\end{equation}
Here the unit vector ${\overline{\bf e}}$ is defined as ${\overline{\bf e}}
\equiv (\overline{\bf p}+\overline{\bf p}^{\, \prime})/
\mid \overline{\bf p}+\overline{\bf p}^{\, \prime} \mid$
($\simeq \overline{\bf p}/\mid \overline{\bf p}\mid$ in the eikonal limit), and 
the overline on the various quantities in the equations above
is to indicate that they are given in the c.m.~frame.  The factor
$(2 {\rm i}\, \overline{p})^{-1}$ in the l.h.s.~of Eq.~(\ref{eq:fnn})
is conventional.

The functions $\overline{F}^{NN}_m$ are parameterized as

\begin{equation}
\overline{F}^{\, NN}_m(s,\overline{\bf k}^{\, 2})= \overline{\alpha}^{NN}_m\!(s) \,\,
{\rm exp}\left[ -\overline{\beta}^{NN}_m\!(s)\, \overline{\bf k}^{\, 2}\right] \ ,
\end{equation}
where the $\overline{\alpha}^{NN}_m\!(s)$ and $\overline{\beta}^{NN}_m\!(s)$
coefficients depend on $s$ and are generally complex; in particular,
the forward, spin-independent amplitude $\overline{F}^{\, NN}_1(s,0)$ is given
by $\overline{F}^{\, NN}_1(s,0)=\sigma^{NN}(1-{\rm i} \rho^{NN})/(8\pi)$ (an invariant
quantity), where $\sigma^{NN}$ is the total cross section and
$\rho^{NN}$ is the ratio of the imaginary to the real part of $\overline{F}^{\, NN}_1(s,0)$.
In the present work these coefficients are taken from Tables III and IV of
Ref.~\cite{Wallace81}: they were obtained by fitting $N$$N$ amplitudes
derived from the phase-shift analysis of the VPI group~\cite{Arndt82}.  The
tabulations are for $s$ values in the range (3.92--5.09) GeV, corresponding
to lab kinetic energies (210--831) MeV, for both $p$$p$ and $p$$n$ scattering.
A detailed assessment of the accuracy and limitations of these parameterizations (of
course, in relation to the $N$$N$ scattering data available up to 1981) can be
found in Ref.~\cite{Wallace81}.  Here, it suffices to note only that they
are reasonably accurate for the central and single spin-flip terms
for $\overline{{\bf k}}^{\,2}$ up to $\simeq 0.1$ (GeV/c)$^2$; however,
double spin-flip terms are not well fitted by (single) Gaussian functions.

It would be desirable to update and improve the parameterizations of
Ref.~\cite{Wallace81} by using phase-shift analyses based on the current
$N$$N$ database.  Unfortunately, at the higher lab energies of interest
here (say, above 500 MeV) the increase in the size of this database
has been rather modest since the early 1980's.

It is worth pointing out that all previous Glauber calculations of
$A$($e,e^\prime p)$ processes we are aware of have ignored the double
spin-flip terms in the scattering amplitude, corresponding to $m$=2, 4,
and 5 in Eq.~(\ref{eq:fnn}); indeed, most have only included the spin
independent term (see, for example, Refs.~\cite{Benhar00,Ryckebusch03,Ciofi05,Ciofi05a}).

Finally, note that, if the amplitude in Eq.~(\ref{eq:fnn}) were
to consist only of the scalar term ($m$=1), then the transformation
to the lab frame would be unnecessary, since
$(2 {\rm i}\, \overline{p})^{-1}\, \overline{F}^{\, NN}_{ij}$ would be
an invariant function of $s$ and the four-momentum transfer squared
$t$=$-\overline{\bf k}^{\, 2}$.  However, the presence of the
spin-dependent terms spoils this simplicity.
\subsection{Boosting the $N$$N$ scattering amplitude to the rescattering frame}
\label{sec:flab}

In order to boost the $N$$N$ scattering amplitude from the c.m.~to the
rescattering frame, we adopt the procedure described in Ref.~\cite{McNeil83},
although its practical implementation in the present work is approximate for reasons
discussed below.  First, an invariant representation of the amplitude is
introduced in terms of scalar, vector ($\gamma^\mu$), tensor ($\sigma^{\mu \nu}$),
pseudoscalar ($\gamma^5$), and axial vector ($\gamma^5\gamma^\mu$)
combinations of Dirac matrices:

\begin{equation}
{\cal F}^{NN}_{ij}=\sum_{m=1}^5 {\cal F}_m^{NN}(s,t) \Lambda_{ij}^m \ ,
\end{equation}
where the five operators $\Lambda_{ij}^m$ are defined as

\begin{equation}
\Lambda_{ij}^{m=1,\dots,5}=1\, ,\, \gamma_i^\mu \gamma_{j,\mu} \, , \,
\sigma_i^{\mu\nu} \sigma_{j,\mu\nu} \, , \, \gamma_i^5 \gamma_j^5 \, , \,
 \gamma^5_i \gamma_i^\mu \gamma^5_j \gamma_{j,\mu}  \ .
\end{equation}
The relation between the functions ${\cal F}^{NN}_m$ and $\overline{F}^{\, NN}_m$
in the c.m.~frame follows by noting that 

\begin{equation}
\overline{u}_{\sigma_i^\prime}( \overline{\bf p}^{\, \prime})
\overline{u}_{\sigma_j^\prime}(-\overline{\bf p}^{\, \prime}) {\cal F}^{NN}_{ij}
u_{\sigma_i}( \overline{\bf p}) 
u_{\sigma_j}(-\overline{\bf p}) =
\chi_{\sigma_i^\prime}^\dagger \chi_{\sigma_j^\prime}^\dagger \left[ 
(2 {\rm i}\, \overline{p})^{-1}\, \overline{F}^{\, NN}_{ij}(\,\overline{{\bf k}},s)\right] 
\chi_{\sigma_i} \chi_{\sigma_j} \ ,
\end{equation}
where the $u_\sigma$ are (positive-energy) Dirac spinors with
$\overline{u}_\sigma \equiv u_\sigma^\dagger \gamma^0$, and
$\chi_\sigma$ are two-component Pauli spinors.  This leads
in the eikonal limit $\overline{\bf p} \simeq \overline{\bf p}^{\, \prime}$ to 

\begin{equation}
{\cal F}^{NN}_m(s,\overline{\bf k}^{\, 2})= \sum_{n=1}^5
\overline{M}_{mn}(\overline{p},\overline{\bf k}^{\, 2})\,
\overline{F}^{\, NN}_n(s,\overline{\bf k}^{\, 2}) \ ,
\label{eq:flab}
\end{equation}
where the matrix $\overline{M}$ is the inverse of that given
in Tables I and II of Ref.~\cite{McNeil83}.  In fact, the $\overline{\bf k}^{\, 2}$
dependence in the matrix $\overline{M}$ is neglected,
i.e.~$\overline{M}\equiv \overline{M}(\overline{p},0)$,
in the present work, which allows one to write ${\cal F}^{NN}_m$, $m=1,\dots,5$,
as a linear combination of terms, each having, as far as the momentum transfer
dependence is concerned, the same Gaussian functional form as $\overline{F}^{\, NN}_n$.
This turns out to be convenient when performing the Fourier transform
in Eq.~(\ref{eq:prof}).

Next, having determined the functions ${\cal F}^{NN}_m$, the
scattering amplitude in the rescattering frame is obtained from 

\begin{equation}
\chi_{\sigma_i^\prime}^\dagger \chi_{\sigma_j^\prime}^\dagger
\left[ (2 {\rm i}\, p)^{-1}\, F^{NN}_{ij}({\bf k},s)\right]
\chi_{\sigma_i} \chi_{\sigma_j}=
\overline{u}_{\sigma_i^\prime}({\bf p}-{\bf k})
\overline{u}_{\sigma_j^\prime}({\bf p}_d/2+{\bf k}) {\cal F}^{NN}_{ij}
u_{\sigma_i}({\bf p}) u_{\sigma_j}({\bf p}_d/2) \ ,
\label{eq:fres}
\end{equation} 
where in practice the dependence upon ${\bf p}_d/2$ in the
spinors of particle $j$ has been neglected (in this limit,
the rescattering and lab frames coincide).  This approximation
is justified at low $p_d$ (corresponding to low missing momenta),
but is clearly unsatisfactory at high $p_d$ ($p_d \simeq 1$ GeV/c
in the experiment of Ref.~\cite{Rvachev05}).  The resulting
$F^{NN}_{ij}({\bf k},s)$ has central, single and double spin-flip
terms.

The approximations above---neglecting the
dependence of the matrix $\overline{M}$
on the momentum transfer in Eq.~(\ref{eq:flab}),
and the momentum ${\bf p}_d/2$ in Eq.~(\ref{eq:fres})--made in boosting
the $N$$N$ amplitude from the c.m.~to the rescattering
frame, have been dictated by computational convenience
rather than by necessity, and could be removed.
This latter task, however, is beyond the scope of the
present work.
\section{Cross section and response functions}
\label{sec:form}

To set the stage for the discussions that follow in later sections,
it is helpful to give the expression of the five-fold differential
cross section for the $^A i$$(e,e^\prime p)$$^{(A-1)}\! f$ process (for
a derivation, see Ref.~\cite{Raskin89}):

\begin{equation}
\frac{ {\rm d}^5 \sigma}{ {\rm d} E_e^\prime {\rm d} \Omega_e^\prime {\rm d} \Omega } =
p\, E\, \sigma_{\rm Mott} f_{\rm rec} \frac{m}{E} \frac{m_f}{E_f}
\left[ v_L R_L + v_T R_T + v_{LT} R_{LT} {\rm cos} (\phi) +
                           v_{TT} R_{TT} {\rm cos}(2 \phi)  \right] \ ,
\label{eq:xxx}
\end{equation}
where $E_e^\prime$ is the energy of the final electron, $\Omega_e^\prime$
and $\Omega$ are the solid angles of, respectively, the final electron
and knocked-out proton, $m_f$ is the rest mass of the ($A$--1)-cluster
(assumed bound here), ${\bf p}$ and $E$ (${\bf p}_f$ and $E_f$) are the
momentum and energy of the proton (($A$--1)-cluster) in the lab system,
$\phi$ is the angle between the electron scattering plane and the plane
defined by the momenta ${\bf q}$ and ${\bf p}$, and the recoil factor
$f_{\rm rec}$, or rather its inverse, is defined as

\begin{equation}
f^{-1}_{\rm rec} = \Bigg| 1 -
\frac{ p_f E} { p E_f} \hat{\bf p} \cdot \hat{\bf p}_f  \Bigg| \ .
\end{equation}
The Mott cross section $\sigma_{\rm Mott}$ and the coefficients $v_{\alpha}$,
$\alpha$=$L$, $T$, $LT$, and $TT$, are defined in terms of the electron kinematical
variables, as given in Eqs.~(2.19) and~(2.27$a$)--(2.27$d$) of Ref.~\cite{Raskin89}
(note, however, that in that work $Q^2$ is taken to be negative).

The response functions $R_\alpha$ involve matrix elements of the charge and
current operators between the initial $^A i$ and final $p$+$^{(A-1)}\! f$
nuclear states, and depend on the momenta $q$ and $p$, the angle $\theta$
between them, and the energy transfer $\omega$.  In a schematic notation,
they are given by

\begin{eqnarray}
R_L &=& |\langle p+ ^{(A-1)}\! f\mid \rho(q\hat{\bf z})\mid ^A i\rangle|^2 \ , \\
R_T &=& |\langle p+ ^{(A-1)}\! f\mid {\bf j}_\perp (q \hat{\bf z})\mid ^A i\rangle|^2 \ , \\
R_{LT} &=& 2\sqrt{2}\, \langle p+  ^{(A-1)}\! f\mid \rho(q \hat{\bf z})\mid ^A i\rangle^*
\, \langle  p+ ^{(A-1)}\! f\mid j_x(q \hat{\bf z})\mid ^A i\rangle \ , \\
R_{TT} &=& -|\langle p+ ^{(A-1)}\! f\mid j_x(q \hat{\bf z})\mid ^A i\rangle|^2 +
|\langle p+ ^{(A-1)}\! f\mid j_y(q \hat{\bf z})\mid ^A i\rangle|^2 \ , 
\end{eqnarray}
where the $z$-axis has been taken along ${\bf q}$, which also defines the
spin-quantization axis, ${\bf j}_\perp$ denotes the components of the current
transverse to ${\bf q}$, and the average over the initial, and sum over the
final, spin projections are understood.

\subsection{The plane-wave-impulse-approximation}
\label{sec:pwia}

In the plane-wave-impulse-approximation (PWIA) limit, in which final-state
interactions (FSI) effects between the knocked-out proton and the nucleons
in the recoiling cluster are ignored (i.e., the operator $G$ in
Eq.~(\ref{eq:glb}) is set to one), the response functions
$R^{\rm PWIA}_\alpha$ can be expressed, neglecting two-body terms in
the electromagnetic current operator, as

\begin{equation}
R^{\rm PWIA}_\alpha =  r^p_\alpha \, N_{pf}(p_{\rm m}) \ ,
\label{eq:rpwia}
\end{equation}
where the $r^p_\alpha$ denote appropriate combinations of kinematical
factors with the proton electric and magnetic form factors---those
corresponding to the one-body currents of Sec.~\ref{sec:cnt} are
listed in Appendix~\ref{app:rp}---and $N_{pf}(p_{\rm m})$
is the $p$+($A$--1)-cluster momentum distribution, defined as

\begin{equation}
N_{pf}(p_{\rm m}) = \frac{1}{2J_i+1} \,\sum_{\sigma_i,\sigma,\sigma_f}
| A^{pf}_{\sigma,\sigma_f;\sigma_i} ({\bf p}_{\rm m}) |^2 \ ,
\end{equation}
with

\begin{equation}
A^{pf}_{\sigma,\sigma_f;\sigma_i} ({\bf p}_{\rm m})=\sqrt{\frac{A}{(2\pi)^3}}\,
\int {\rm d}{\bf r}_1 \dots
{\rm d}{\bf r}_A\, {\rm e}^{-{\rm i} {\bf p}_{\rm m} \cdot
( {\bf r}_A -{\bf R}_{1 \dots A-1} )}\, \chi^\dagger_\sigma(A;p) \,
\phi^\dagger_{\sigma_f}(1 \dots A-1;f) \, \psi_{\sigma_i}(1 \dots A;i) \ .
\end{equation}
Here $J_i$ is the total angular momentum of the initial state, and ${\bf p}_{\rm m}$
is the so-called missing momentum, ${\bf p}_{\rm m}$=$-{\bf p}_f$=${\bf p}-{\bf q}$.
The normalization integral

\begin{equation}
N_f=4\pi \int_0^\infty {\rm d}p_{\rm m}\, p_{\rm m}^2 \,N_{pf}(p_{\rm m}) \ 
\end{equation}
gives the number of $^{(A-1)}\, f$ clusters in the ground
state $^A i$~\cite{Schiavilla86}.  In $^3$He the number of deuterons
is calculated to be about 1.34 (see below), which implies that in $^3$He
(to the extent that it is a pure total isospin $T$=1/2 state), out of a
possible number of 1.5 pairs of nucleons in isospin $T$=0 states, roughly 90\%
of them are in the deuteron state.  Similarly, in $^4$He the number of tritons
is found to be $\simeq 1.68$, and so in $^4$He (again, ignoring admixtures 
of states with $T > 0$, induced by small isospin symmetry-breaking
interactions), about 85\% of the $n$$n$$p$ clusters are in the triton
state~\cite{Schiavilla86,Forest96,Schiavilla05}.

\begin{figure}[bthp]
\includegraphics[width=6in]{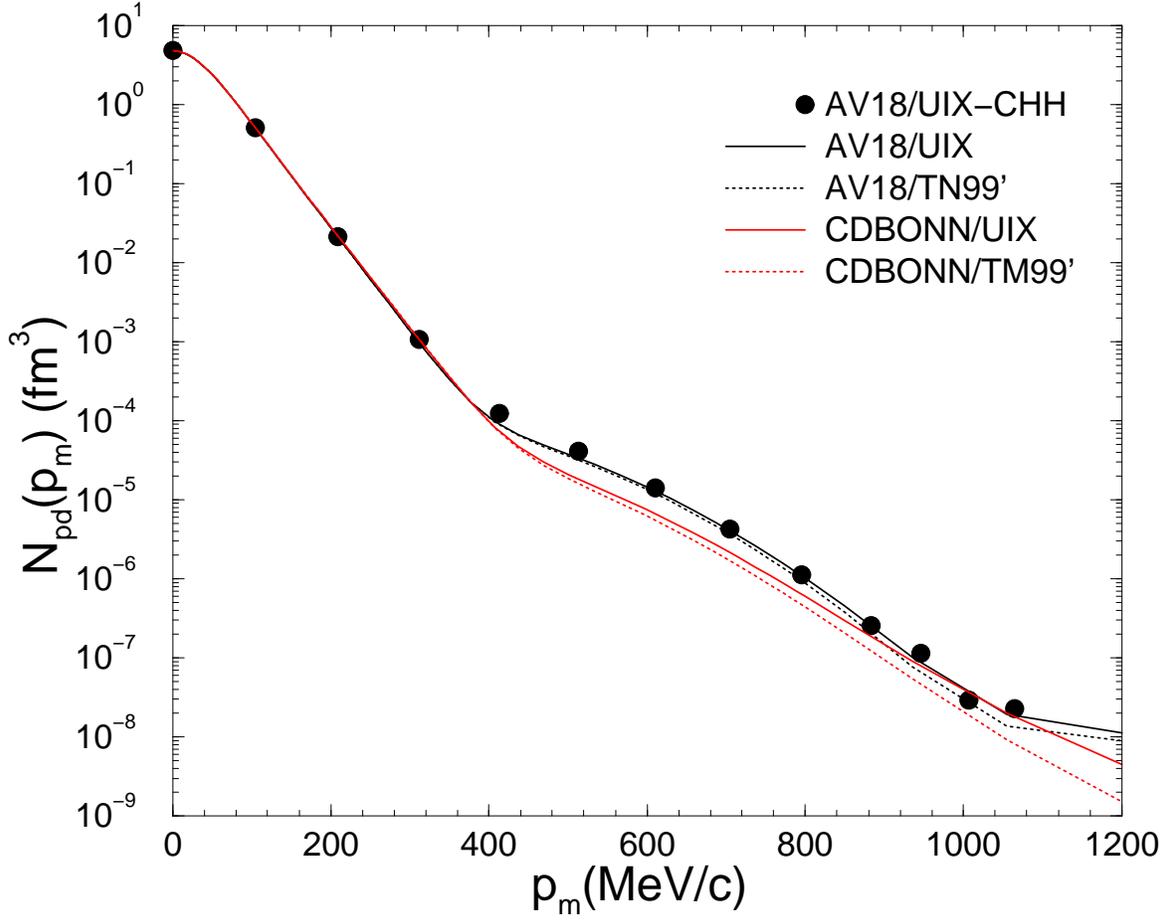}
\caption{The $p$$d$ momentum distribution, obtained with a correlated-hyperspherical-harmonics
(CHH) wave function corresponding to the AV18/UIX Hamiltonian, is compared to those
obtained with Faddeev wave functions corresponding to different combinations of two- and
three-nucleon interactions.}
\label{fig:npd}
\end{figure}

The $N_{pd}(p_{\rm m})$ momentum distribution, obtained with CHH
wave functions corresponding to the AV18/UIX Hamiltonian
model, is shown in Fig.~\ref{fig:npd} up to missing
momenta of $\simeq 1$ GeV/c.  A number of realistic interactions are currently
available, such as, for example, the CD-Bonn~\cite{Machleidt01} or
Nijmegen~\cite{Stoks94} $N$$N$ and Tucson-Melbourne~\cite{Coon79}
$N$$N$$N$ interactions, and therefore the question arises of how sensitive
to the input Hamiltonian are the high momentum components of this momentum
distribution.  This issue is especially relevant here, since the kinematics
of the JLab $(e,e^\prime p)$ experiments cover a broad range of $p_{\rm m}$
values, as high as 1.1 GeV/c.  It is addressed in Fig.~\ref{fig:npd},
where $p$$d$ momentum distributions, obtained with various combinations
of two- and three-nucleon interactions, are compared with each other
up to $p_{\rm m} \simeq 1$ GeV/c.  All results, but for those labeled
AV18/UIX-CHH, are obtained with Faddeev wave functions~\cite{Nogga04}. 

A couple of comments are now in order.  First, in the $p_{\rm m}$ range
(400--800) MeV/c, there is a significant model dependence: the CD-Bonn
$N_{pd}(p_{\rm m})$ is about a factor of 2 smaller than the AV18 one.
This is likely a consequence of the fact that the tensor force is
weaker in the CD-Bonn than in the AV18.  The $p$$d$ overlap in $^3$He has
S- and D-state components, and the associated D-state contribution to
$N_{pd}(p_{\rm m})$ indeed becomes dominant at $p_m \simeq 400$ MeV/c,
it is responsible for the change of slope in $N_{pd}(p_{\rm m})$.

Second, the Faddeev and CHH wave functions corresponding to the
AV18/UIX Hamiltonian model lead to $p$$d$ momentum distributions,
that are slightly different only in the $p_{\rm m}$ region around
400 MeV/c, where the S-wave contribution changes sign.

\section{Calculation}
\label{sec:calc}

Nuclear wave functions, for an assigned spatial configuration
${\bf R}=({\bf r}_1,\dots,{\bf r}_A)$, are expanded on a basis of
$K$=$2^A A!/[Z! (A-Z)!]$ spin-isospin states for $A$ nucleons
($Z$ is the number of protons) as

\begin{equation}
\psi({\bf R}) = \sum_{k=1}^K \psi_k({\bf R}) \mid\! k\rangle \ ,
\end{equation}
where the components $\psi_k({\bf R})$ are generally complex functions
of ${\bf R}$, and, in the case $A$=3 and $Z$=2 as an example, the basis states $\mid\! k\rangle$=
$\mid\! (n\! \downarrow)_1 (p\! \downarrow)_2 (p\! \downarrow)_3\rangle$,
$\mid\! (n\! \uparrow)_1 (p\! \downarrow)_2 (p\! \downarrow)_3\rangle$, $\dots\ ,$
$\mid\! (p\! \uparrow)_1, (p\! \downarrow)_2 (n\! \downarrow)_3 \rangle$, $\dots \ .$
Matrix elements of the electromagnetic current operator are written schematically as

\begin{equation}
\langle f\mid O \mid i \rangle=\sum_{k,l=1}^K \int {\rm d}{\bf R}\, \psi^*_k({\bf R};f)
O_{kl}({\bf R}) \psi_l({\bf R};i) \ ,
\end{equation}
where $\left[ O_{kl}({\bf R})\right]$ denotes the matrix representing in configuration space
any of the one- or two-body charge/current operators.  Matrix multiplications
in the spin-isospin space are performed exactly with the techniques developed in
Ref.~\cite{Schiavilla89}.  The problem is reduced to the evaluation of the
spatial integrals, which is efficiently carried out with Monte Carlo (MC) methods,
although these are implemented differently in the present study than they have
been in the past~\cite{Schiavilla86,Schiavilla90}.

To illustrate these methods, consider the PWIA calculation of the one-body
charge operator in the process $^3$He($e,e^\prime p$)$d$:

\begin{equation}
\rho^{\rm PWIA} \sim \sum_{k,l} \int{\rm d}{\bf R}\,  
{\rm e}^{-{\rm i} {\bf p}  \cdot {\bf r}_3 } \,
{\rm e}^{-{\rm i} {\bf p}_f \cdot {\bf R}_{12} }\, \psi^*_k({\bf r}_1-{\bf r}_2;f) \,
{\rm e}^{ {\rm i} {\bf q}  \cdot {\bf r}_3 } \, \rho_{kl}({\bf q};-{\rm i} \nabla_3) 
\psi_l({\bf R};i)  \ ,
\end{equation}
where $\psi(i)$ and $\psi(f)$ denote the $^3$He and $p$+$d$ cluster
wave functions, respectively, and $\rho_{kl}({\bf q};-{\rm i} \nabla_3)$
is $k$$l$ matrix element of the charge operator in Eq.~(\ref{eq:rho1}).  Its 
dependence on the momentum operator, through the spin-orbit term, has been
made explicit.  It is convenient to introduce the Jacobi variables,
${\bf x}$=${\bf r}_3-{\bf R}_{12}$, ${\bf y}$=${\bf r}_1-{\bf r}_2$,
so that the integral reads

\begin{equation}
\rho^{\rm PWIA} \sim \sum_{k,l} \int{\rm d} {\bf x} {\rm d} {\bf y} \,
{\rm e}^{-{\rm i} {\bf p}_{\rm m}  \cdot {\bf x} } \,
\psi^*_k({\bf y};f) \rho_{kl}\left({\bf q};-{\rm i} \frac{3}{2}\nabla_{\bf x}\right)
\psi_l({\bf x},{\bf y};i) \ ,
\end{equation}
where ${\bf p}_{\rm m}$ is the missing momentum.  In
Refs.~\cite{Schiavilla86,Schiavilla90}, this integral was
evaluated with the Metropolis algorithm~\cite{Metropolis53} by sampling the
coordinates (${\bf x},{\bf y}$) from a probability density, taken
to consist of products of central correlation functions from VMC wave
functions (see Ref.~\cite{Schiavilla86}).  While this procedure is
satisfactory at low $p_{\rm m}$, in the sense that statistical errors
are small, it becomes impractical as $p_{\rm m}$ increases, due to the
rapidly oscillating nature of the integrand.  Indeed, this problem is evident
in the VMC calculation of the $p$$d$ and $p$$t$ momentum distributions, shown in
Ref.~\cite{Forest96}: even though the random walk consists of a 
large number (of the order of several hundreds of thousands) of configurations,
oscillations in the central values persist at large $p_{\rm m}$ ($ > 600$ MeV/c).

In the present work, we carry out the multidimensional integrations by a combination
of MC and standard quadratures, namely we write 

\begin{equation}
\rho^{\rm PWIA} \sim \int{\rm d} {\bf y} \int_0^{2\pi} {\rm d}\phi_x
W(\phi_x, {\bf y}) \, F(\phi_x, {\bf y}) \simeq \frac{1}{N_c} \sum_{c=1}^{N_c} F(c) \ ,
\end{equation}
where the $c$'s denote configurations $(\phi_x,{\bf y})$ (total number $N_c$),
sampled with the Metropolis algorithm from a probability density $W$ (normalized to one),
given by

\begin{equation}
W(\phi_x, {\bf y}) = \frac{1}{2\pi} \sum_{k} | \psi_k({\bf y};f) |^2 \ .
\end{equation}
Note that $W$ is uniform in $0 \leq \phi_x \leq 2\pi$.  For each configuration $c$, the function $F$
is obtained by Gaussian integrations over the $x$ and ${\rm cos}\, \theta_x$ variables, i.e.

\begin{equation}
F(c)= \frac{1}{W(c)} \sum_{k,l} \int_{-1}^1 {\rm d}({\rm cos}\theta_x)
\int_0^\infty {\rm d}x\, x^2\, {\rm e}^{-{\rm i} {\bf p}_{\rm m}  \cdot {\bf x} } \, 
\psi^*_k({\bf y};f) \rho_{kl}\left({\bf q};-{\rm i} \frac{3}{2}\nabla_{\bf x}\right)
\psi_l({\bf x},{\bf y};i) \ .
\end{equation}
As a result, the statistical errors are very significantly reduced.  In Fig.~\ref{fig:npd}
the CHH calculation of $N_{pd}(p_{\rm m})$ is carried out with this method, it uses a random walk
consisting only of 20,000 configurations.  However, convergence in the ($x$,${\rm cos}\,\theta_x$)
integrations requires of the order of (70,50) Gaussian points at the highest $p_{\rm m}$, and so
the present method turns out to be computationally more time-consuming than the earlier version at
high $p_{\rm m}$.

Additional refinements in the present MC implementation are i) the application of
gradient operators on the left, rather than right, wave function, and ii) the use
of block averaging for a more realistic estimation of the statistical errors.

Gradient operators, such as those occurring in the one-body electromagnetic
current, are discretized as

\begin{equation}
\nabla_{i,\alpha} \psi({\bf R})\simeq \frac{\psi(\dots {\bf r}_{i}+\delta\, \hat{\bf e}_\alpha \dots)
-\psi(\dots {\bf r}_{i}-\delta\, \hat{\bf e}_\alpha \dots)}{2\,\delta} \ ,
\end{equation}
where $\delta$ is a small increment ($\delta$=0.0005 fm in the calculations reported here) and
$\hat{\bf e}_\alpha$ is a unit vector in the $\alpha$-direction.  Therefore, again in the
context of the PWIA calculation above, the $\nabla_{\bf x}\ ,$ when
operating on the left, only acts on the plane wave---in fact, an eigenfunction of $\nabla_{\bf x}\ .$
This further reduces statistical errors, and also ensures that the PWIA relations in Eq.~(\ref{eq:rpwia})
are satisfied---modulo tiny discretization errors of order $(\delta\, p_{\rm m})^2$---at each configuration
in the random walk, which would not be the case if the gradient were left to operate
on the $^3$He wave function to the right.  Of course, the eigenfunction property above
is spoiled, when final state interactions are taken into account; however, the error-reduction
benefits remain.  The disadvantage of the procedure just outlined is that it leads
to an increase in computational time, since the various gradients have to be evaluated, rather
than once (when acting to the right), as many times as the number of kinematics being considered
in the calculation.

A crude estimate of the MC error is obtained as

\begin{equation}
\Delta(F) = \frac{1}{\sqrt{N_c}} \left[ \frac{1}{N_c} \sum_{c=1}^{N_c} F^{\, 2}(c)
   -\left[  \frac{1}{N_c} \sum_{c=1}^{N_c} F(c) \right]^{\! 2}\, \right]^{1/2}  \ ,
\end{equation}
it assumes that the distribution $F(c)$ is Gaussian, whereas in practice this is generally
not the case.  A better estimate, adopted in the present work, is obtained by dividing
the set of $N_c$ samples into $M_c$ blocks containing $n_c$ samples each:

\begin{equation}
f_m = \frac{1}{n_c} \sum_{c=(m-1)\, n_c+1}^{m\, n_c} F(c) \ . 
\end{equation}
Then,

\begin{equation}
\Delta(f) = \frac{1}{\sqrt{M_c}} \left[ \frac{1}{M_c} \sum_{m=1}^{M_c} f^{\, 2}_m
   -\left(  \frac{1}{M_c} \sum_{m=1}^{M_c} f_m \right)^{\! 2}\, \right]^{1/2}  \ ,
\end{equation}
where $n_c$ can be chosen large enough so as to make the distribution of $f_m$ Gaussian (in
practice, $n_c$ has been taken of the order of 100).
\section{Results}
\label{sec:res}

\begin{figure}[bthp]
\includegraphics[width=6in]{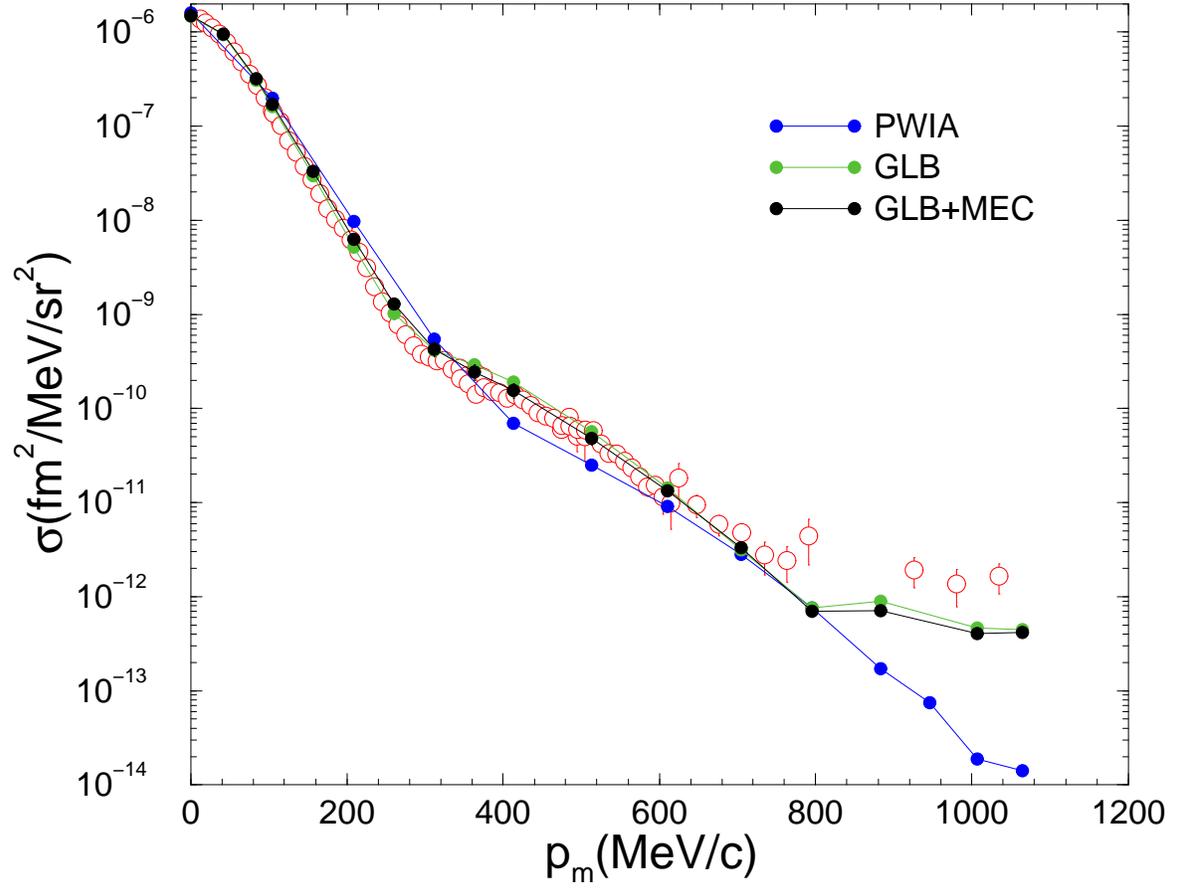}
\caption{The experimental data for the $^3$He$(e,e^\prime p)$$d$ cross section at
$\phi$=180 deg are compared to the results of calculations in plane-wave-impulse-approximation
(PWIA), or using the Glauber approximation without (GLB) and with (GLB+MEC) inclusion of
two-body currents.  The profile operator in the Glauber expansion is derived from the full
$N$$N$ scattering amplitude, boosted from the c.m.~to the the rescattering (i.e., lab) frame.}
\label{fig:xkin1}
\end{figure}

\begin{figure}[bthp]
\includegraphics[width=6in]{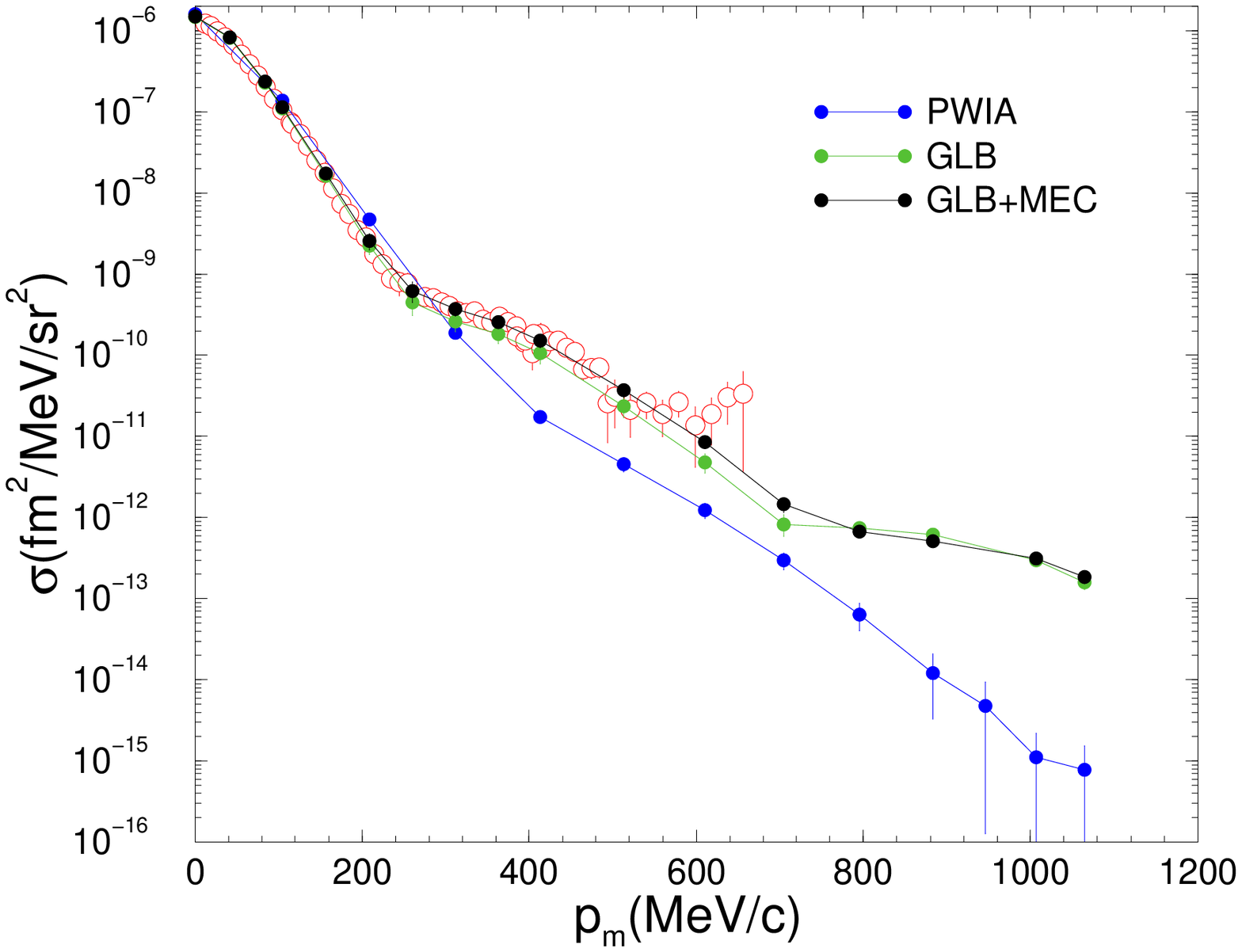}
\caption{Same as in Fig.~\protect\ref{fig:xkin1}, but at $\phi$=0 deg.}
\label{fig:xkin2}
\end{figure}

\begin{figure}[bthp]
\includegraphics[width=6in]{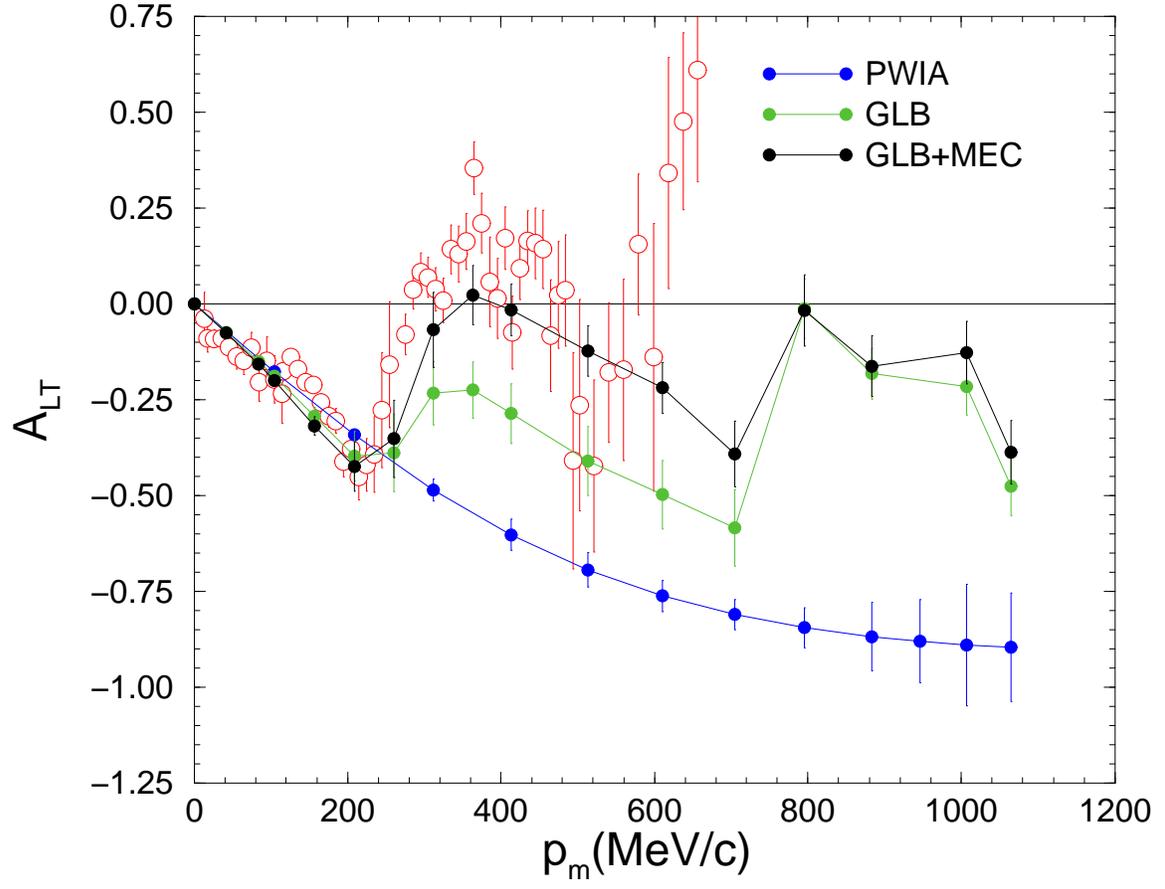}
\caption{Same as in Fig.~\protect\ref{fig:xkin1}, but for the longitudinal-transverse
asymmetry rather than the cross section.}
\label{fig:asy}
\end{figure}

The predicted cross sections are compared with the data taken at
JLab (E89-044)~\cite{Rvachev05} in Figs.~\ref{fig:xkin1} and~\ref{fig:xkin2}.
The in-plane measurements were carried out in quasi-elastic kinematics with
the proton being detected on either side of the three-momentum transfer ${\bf q}$
(the kinematics in Figs.~\ref{fig:xkin1} and ~\ref{fig:xkin2} have,
respectively, $\phi$=180 deg and 0 deg, in the notation of
Sec.~\ref{sec:form}).  From these cross sections, the longitudinal-transverse
asymmetry $A_{LT}$ is obtained as

\begin{eqnarray}
A_{LT}&=&\frac{\sigma(\phi=0\, {\rm deg}) - \sigma(\phi=180\, {\rm deg})}
              {\sigma(\phi=0\, {\rm deg}) + \sigma(\phi=180\, {\rm deg})} \nonumber \\
      &=&\frac{v_{LT}\, R_{LT}}{v_L\, R_L+v_T\, R_T+v_{TT}\, R_{TT}} \ ,
\end{eqnarray}
where $\sigma(\phi)$ stands for the five-fold differential cross
section in Eq.~(\ref{eq:xxx}).  Thus, the asymmetry, shown in
Fig.~\ref{fig:asy}, is proportional to the $R_{LT}$ response
function.  Note that $v_{LT}$ is negative, as defined in
Ref.~\cite{Raskin89}.

In the figures, the results of three different calculations are
displayed.  The curve labeled PWIA is a plane-wave-impulse-approximation
calculation, based on the CHH $^3$He wave function corresponding to the AV18/UIX
Hamiltonian model (the resulting $p$$d$ momentum distribution is shown in
Fig.~\ref{fig:npd}).  It neglects final-state-interaction (FSI) effects and
contributions from two-body currents (MEC).  The present PWIA results are
in agreement with those of recent studies~\cite{Ciofi05a,Laget05}: they overpredict
the measured cross section for values of the missing momentum $p_{\rm m}$ up to
$\simeq 300$ MeV/c, while underpredicting it at high $p_{\rm m}$.  This underprediction
is particularly severe for the kinematics with $\phi$=0 deg (Fig.~\ref{fig:xkin2}):
the contribution proportional to $R_{LT}$ in Eq.~(\ref{eq:xxx}) is comparable in
magnitude for $p_{\rm m} > 300$ MeV/c, but of opposite sign, relative to that from $R_L$
and $R_T$ (the $R_{TT}$ contribution is negligible).  Of course, when
$\phi$=180 deg, they have the same sign.

The curve labeled GLB (GLB+MEC) shows the results of calculations using
the Glauber approximation to describe FSI and one-body (one- plus two-body)
currents.  Both single- and double-rescattering terms are retained in the
Glauber treatment of the final $p$+$d$ scattering state.  The profile
operator is obtained from the full $N$$N$ scattering amplitude, boosted
from the c.m.~to the rescattering frame as discussed in Sec.~\ref{sec:flab}.  The
invariant energy $\sqrt{s}$ at which the rescattering process occurs,
Eq.~(\ref{eq:kin}), changes little over the whole range of missing momenta,
it corresponds to a lab kinetic energy of $\simeq$ 830 MeV.  Thus the parameters
used for the central, single and double spin-flip terms of the
$p$$p$ and $p$$n$ amplitudes are taken from the last row of, respectively,
Tables III and IV of Ref.~\cite{Wallace81}.  However, the parameters
for the $B^{pn}(k)$ and $E^{pn}(k)$ terms (in the notation
of Ref.~\cite{Wallace81}), corresponding to $m$=2 and 5 in
Eq.~(\ref{eq:fnn}), are replaced by those for $B^{pp}(k)$ and $E^{pp}(k)$,
since the real parts of $\beta_B^{pn}$ and $\beta_E^{pn}$ are negative.

There is satisfactory agreement between theory and experiment up to
missing momenta of 700 MeV/c.  At values of $p_{\rm m} \simeq 1$ GeV/c, however,
the theoretical results are smaller than the experimental values
by about a factor of two, although they do reproduce the flattening of
the cross section, as function of $p_{\rm m}$, seen in the data.  Final-state-interaction
effects play a crucial role: they reduce the PWIA cross section at low $p_{\rm m}$
($< 300$ MeV/c), and increase it very substantially at larger values of $p_{\rm m}$.
Two-body current contributions, while not large, are not negligible, and
improve the agreement between theory and experiment.  This is most clearly
illustrated in the case of the $A_{LT}$ observable, Fig.~\ref{fig:asy}.

It would be interesting to investigate the model dependence due to
the input Hamiltonian used to generate the bound-state wave functions.
We do not expect it to be large, even for $p_{\rm m} =(400$--800) MeV/c
where the calculated $p$$d$ momentum distributions can differ by as much as
a factor of two, see Fig.~\ref{fig:npd}.  The cross section in this range
of $p_{\rm m}$ values results from strength shifted by FSI from the low $p_{\rm m}$ 
region, $p_{\rm m}$ up to $\simeq 300$ MeV/c, where the $N_{pd}(p_{\rm m})$ model
dependence is negligible.  Clearly, a direct calculation is needed to verify
whether this expectation is justified.

\begin{figure}[bthp]
\includegraphics[width=6in]{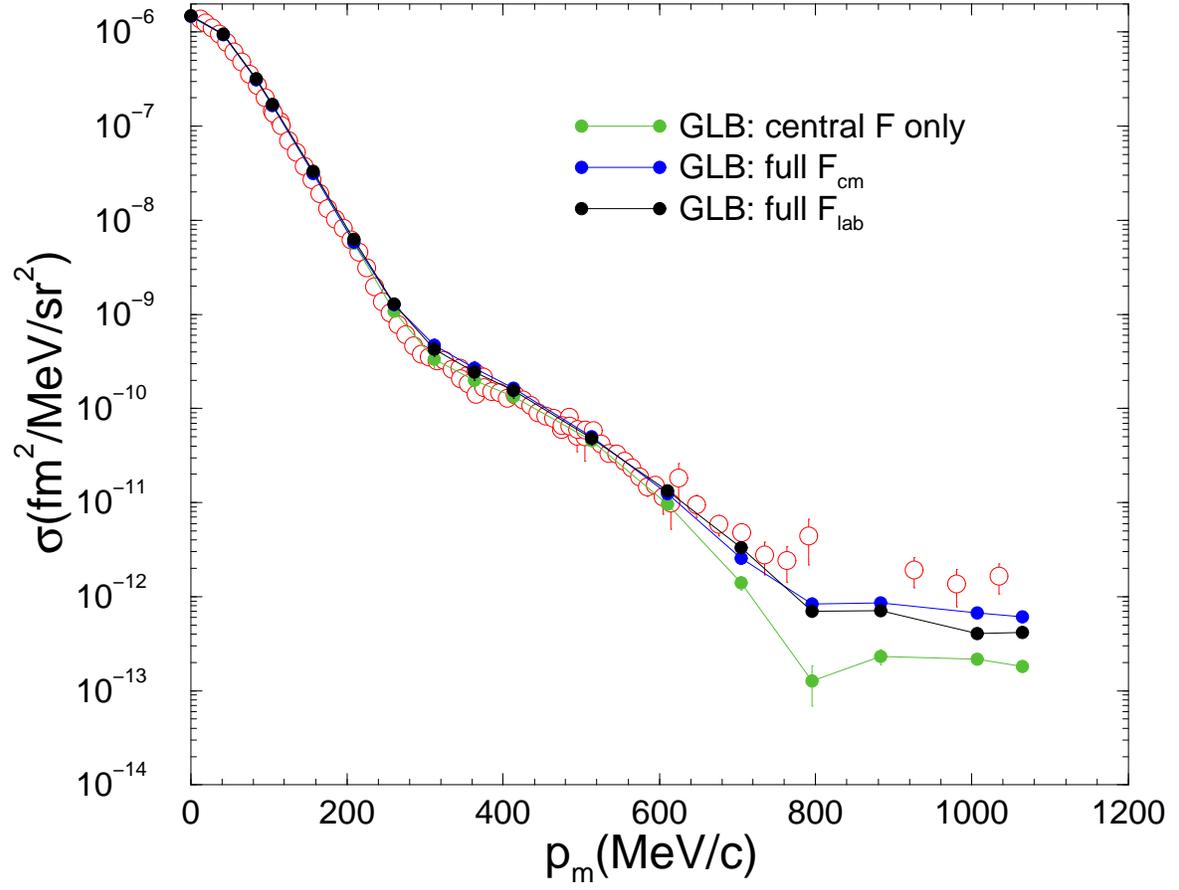}
\caption{The experimental data for the $^3$He$(e,e^\prime p)$$d$ cross section at
$\phi$=180 deg are compared to results of Glauber calculations, using a variety
of approximation schemes for the $N$$N$ scattering amplitude.  See text for
an explanation of the notation.}
\label{fig:xcmp1}
\end{figure}

\begin{figure}[bthp]
\includegraphics[width=6in]{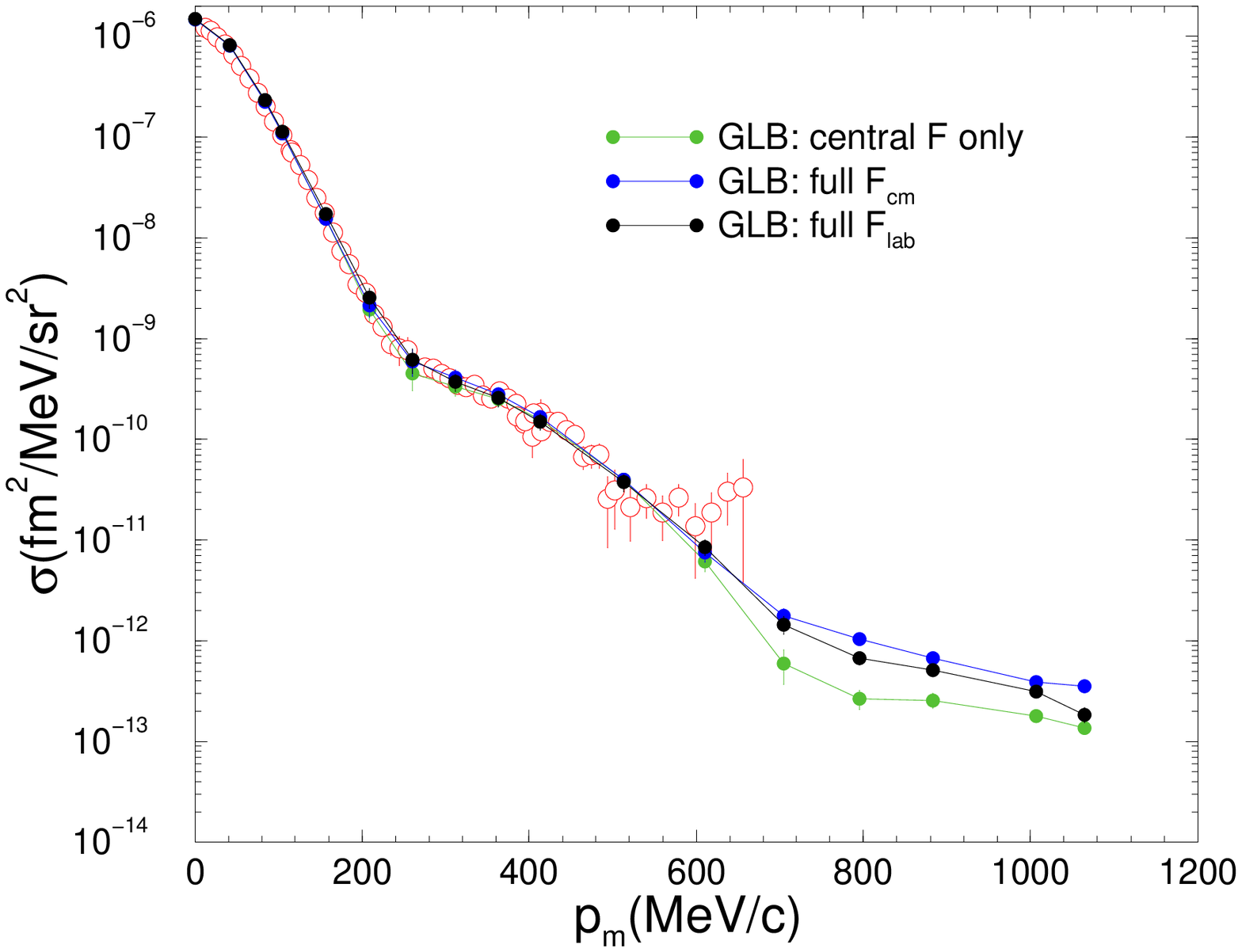}
\caption{Same as in Fig.~\protect\ref{fig:xcmp1}, but at $\phi$=0 deg.}
\label{fig:xcmp2}
\end{figure}

\begin{figure}[bthp]
\includegraphics[width=6in]{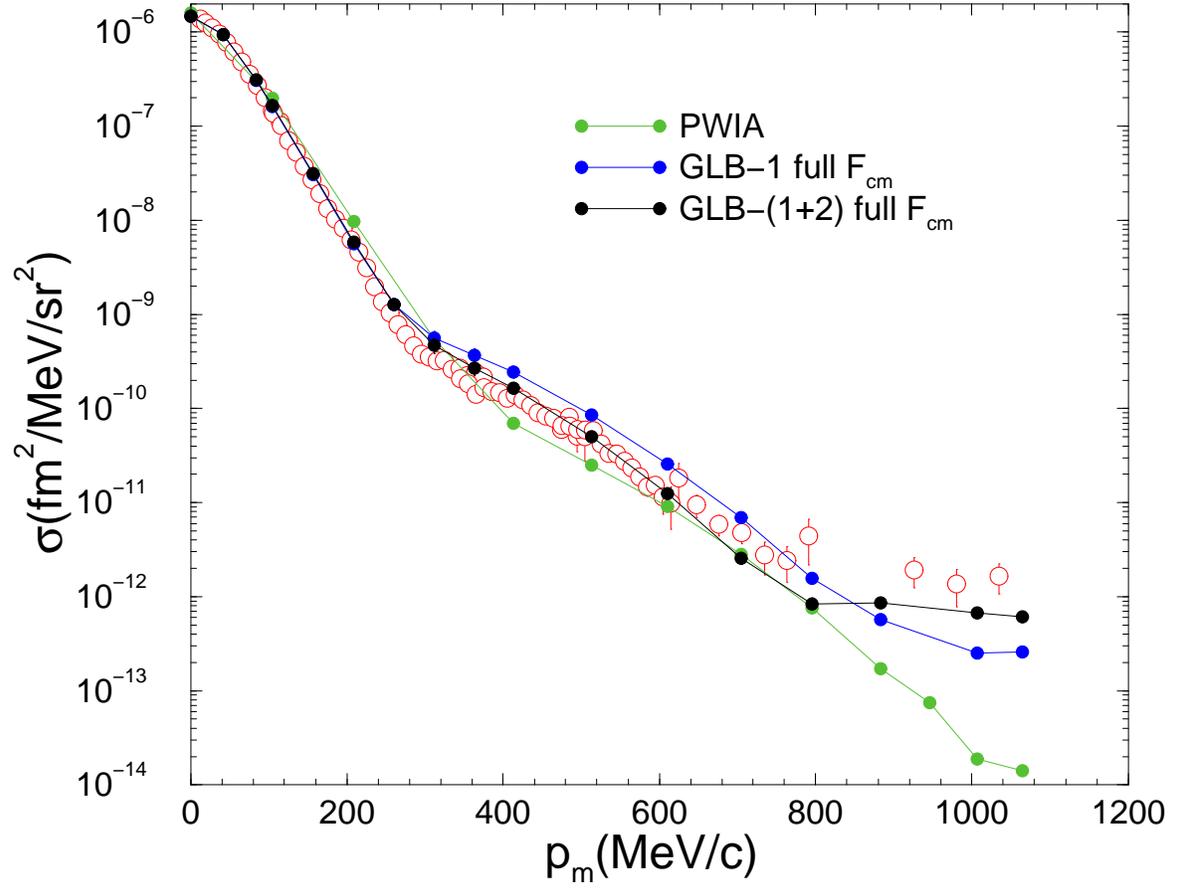}
\caption{The experimental data for the $^3$He$(e,e^\prime p)$$d$ cross section at
$\phi$=180 deg are compared to Glauber calculations including only single (GLB-1)
or both single and double [GLB(1+2)] rescattering terms.  Boost corrections in
the $N$$N$ scattering amplitude are neglected.  Note that two-body current contributions
are included.}
\label{fig:xcmp1a}
\end{figure}

\begin{figure}[bthp]
\includegraphics[width=6in]{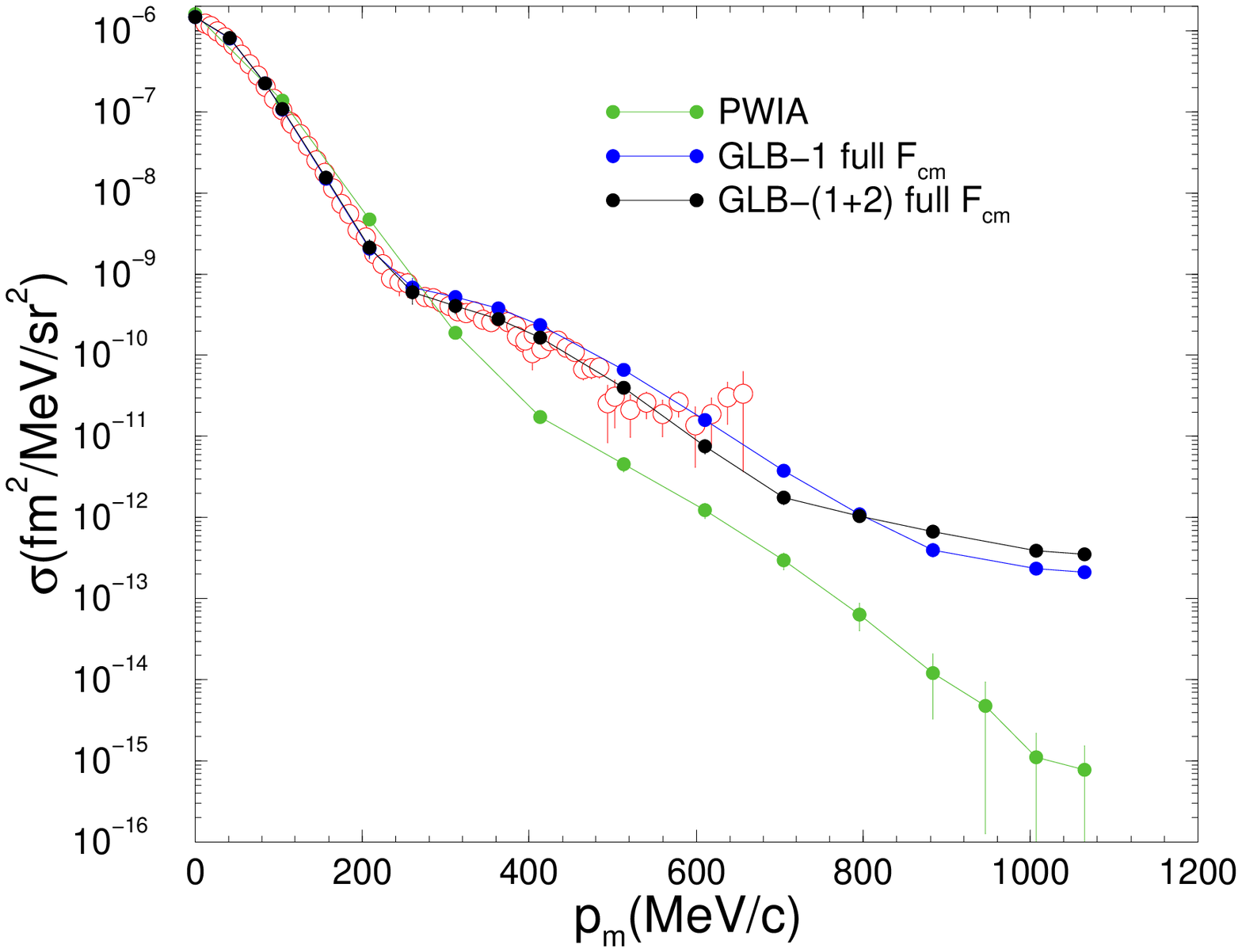}
\caption{Same as in Fig.~\protect\ref{fig:xcmp1a}, but at $\phi$=0 deg.}
\label{fig:xcmp2a}
\end{figure}

\begin{figure}[bthp]
\includegraphics[width=6in]{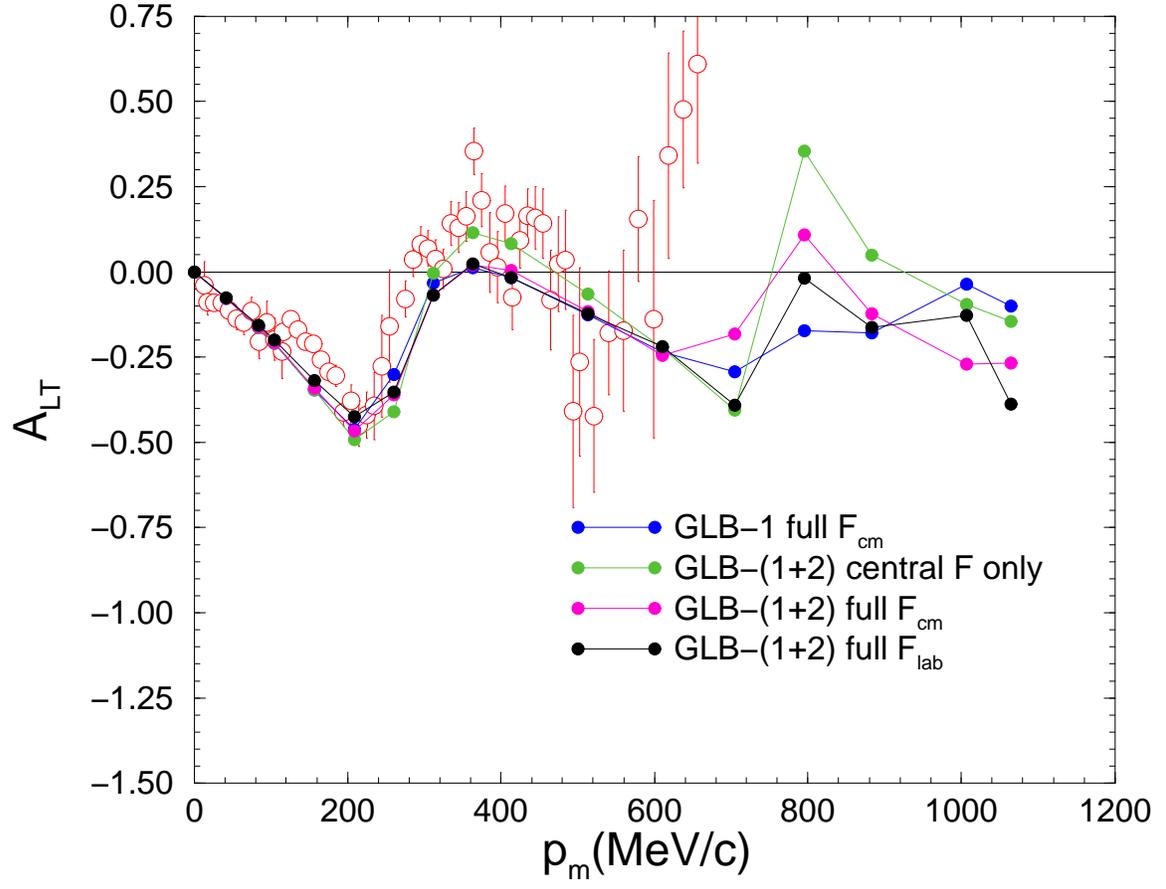}
\caption{The experimental data for the longitudinal-transverse asymmetry
in the $^3$He$(e,e^\prime p)$$d$ reaction are compared to the results of
Glauber calculations including only single (GLB-1) or both single and
double [GLB-(1+2)] rescattering terms.  The GLB-(1+2) calculations use
the full $N$$N$ scattering amplitude, with (full $F_{\rm lab}$) or without
(full $F_{\rm cm}$) boost corrections, or its central term only (central $F$ only).
The GLB-1 calculation uses the full $N$$N$ scattering amplitude, but ignores
boost effects.  Note that two-body current contributions are included
in all calculations.}
\label{fig:asycmp}
\end{figure}

The next set of figures, Figs.~\ref{fig:xcmp1}--\ref{fig:asycmp}, is meant to
illustrate the effect of various approximation schemes in the Glauber treatment
of FSI.  In all calculations MEC contributions are included.  In Figs.~\ref{fig:xcmp1}
and~\ref{fig:xcmp2} the results of calculations using i) only the central part of the
scattering amplitude (curves labeled \lq\lq central F only\rq\rq) and ii) the full
scattering amplitude but neglecting boost corrections (curves labeled \lq\lq full
F$_{\rm cm}$\rq\rq) are compared with the baseline GLB+MEC results of Figs.~\ref{fig:xkin1}
and~\ref{fig:xkin2} (labeled here as \lq\lq full F$_{\rm lab}$\rq\rq).  For values
of $p_{\rm m} > 600$ MeV/c, the spin-dependence of the $N$$N$ scattering amplitude,
which is ignored in all Glauber calculations of $(e,e^\prime p)$ reactions
off few-body nuclei we are aware of (see, for example,
Refs.~\cite{Benhar00,Ryckebusch03,Ciofi05,Ciofi05a}), leads to a very substantial
increase of the cross section obtained when using the central term only.  On the other
hand, boost corrections, which in the present work are only accounted for approximately
(see Sec.~\ref{sec:flab}), seem to be small, although not negligible. 
 
In Figs.~\ref{fig:xcmp1a} and~\ref{fig:xcmp2a}, the results of calculations using
the full scattering amplitude in the c.m.~frame but including only single
rescattering in the Glauber expansion (curve
labeled \lq\lq GLB-1 full F$_{\rm cm}$\rq\rq), i.e.~the term $G^{(1)}$ of Eq.~(\ref{eq:glb1}), 
are compared with the predictions obtained by retaining both single and double
rescatterings (the curve labeled \lq\lq GLB-(1+2) full F$_{\rm cm}$\rq\rq is
the same as \lq\lq full F$_{\rm cm}$\rq\rq in the previous two figures).
Also shown are the PWIA results.  At low missing momenta (below 300 MeV/c)
interference between the plane-wave and single-rescattering amplitudes 
reduces the PWIA cross section.  In the $p_{\rm m}$ range $\simeq (300-800)$ MeV/c
destructive interference occurs between the leading single- and double-rescattering
amplitudes, resulting in a reduction of the cross section obtained in the GLB-1 calculation.
At the highest values of $p_{\rm m}$, double-rescattering processes are dominant. 
The interference pattern among these various amplitudes is consistent with that
obtained in the calculations of Ref.~\cite{Ciofi05a}.

Lastly, the $A_{LT}$ asymmetry is found to be relatively insensitive to the
various approximation schemes discussed above in the region where measurements
are available ($p_{\rm m} < 600$ MeV/c).  All calculations reproduce quite well
the oscillating behavior of the $A_{LT}$ data.

As already mentioned, most Glauber calculations of $A(e,e^\prime p)$ reactions
have only used the central part of the $N$$N$ scattering amplitude $F_{ij}^{NN}$, i.e.

\begin{equation}
(2 {\rm i}\, p)^{-1}\, F^{NN}_{ij}({\bf k},s) \longrightarrow 
\frac{1}{8\pi} \sigma^{NN}(s)\left[ 1-{\rm i} \rho^{NN}(s)\right]\,
{\rm exp}\left[-\beta^{NN}(s) {\bf k}^2 \right] \ ,
\end{equation}
where $\sigma^{NN}(s)$ and $\rho^{NN}(s)$ are, respectively, the total cross
section and ratio of the real to imaginary part of the scattering amplitude
at the invariant energy $\sqrt{s}$.
While the values for $\sigma^{NN}(s)$ and, to a less extent, $\rho^{NN}(s)$ are
well known for both $p$$p$ and $p$$n$ over a wide range of $\sqrt{s}$ (see, for example,
Ref.~\cite{Ryckebusch03} and references therein), this is not the case for $\beta^{NN}(s)$.
This parameter is determined by fitting either the elastic differential cross section
at forward scattering angles (as in Ref.~\cite{Ryckebusch03} and references therein),
i.e.~${\rm d}\sigma^{NN}_{\rm el}(s,t)/{\rm d}t$ for small four-momentum transfer $t$,
or rather the central term of the scattering amplitude derived from phase-shift analyses
(as in Ref.~\cite{Wallace81} and references therein).  It should be emphasized
that the first procedure tacitly assumes that the contributions to
${\rm d}\sigma^{NN}_{\rm el}(s,t)/{\rm d}t$ due to spin-dependent terms in $F_{ij}^{NN}$
are negligible.  It is not obvious that this assumption is justified.  For
example, large cross section differences, $\Delta \sigma_L^{NN}$, between spin
orientations parallel and antiparallel to the beam direction are observed in $p$$p$
and $p$$n$ scattering~\cite{Lechanoine93}.  Also observed are substantial, although
less dramatic, cross section differences, $\Delta \sigma_T^{NN}$, between parallel
and antiparallel transverse spin orientations~\cite{Lechanoine93}.
 
\begin{figure}[bthp]
\includegraphics[width=6in]{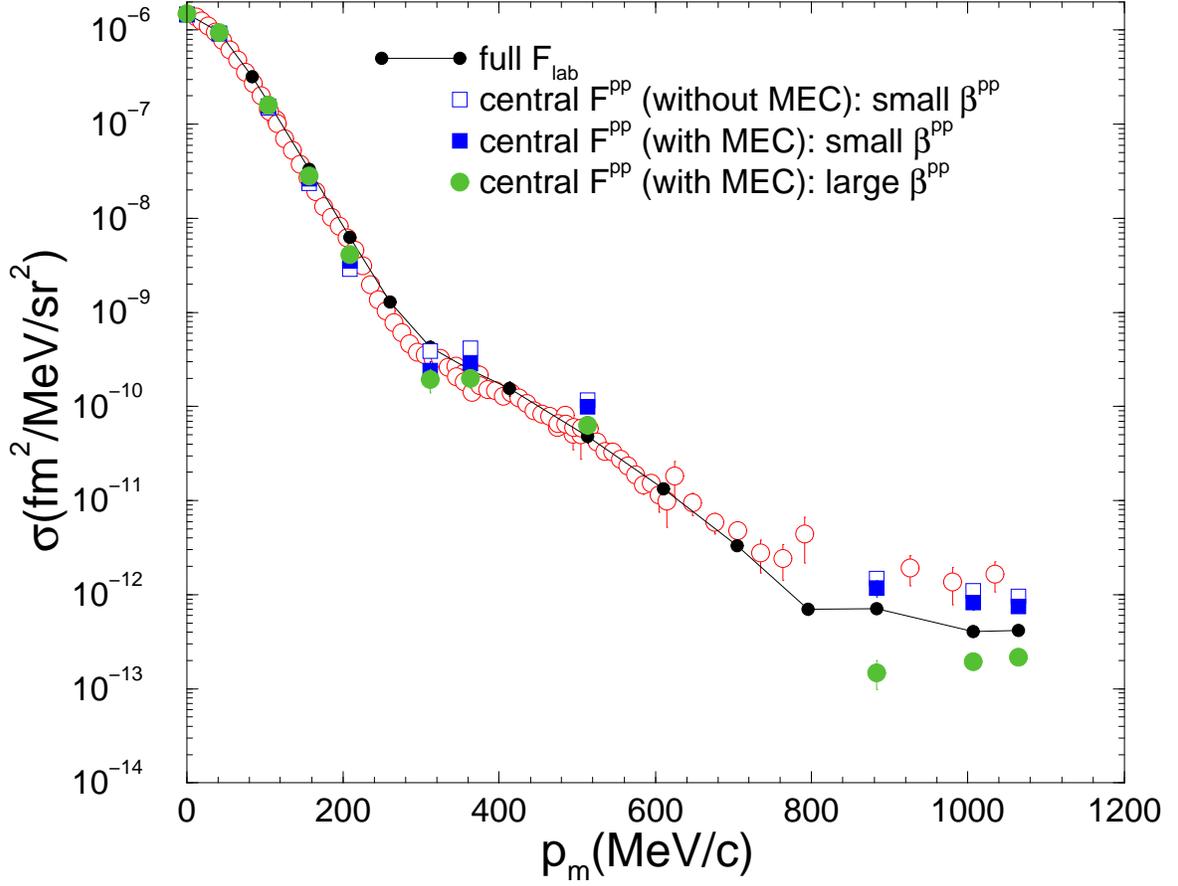}
\caption{The experimental data for the $^3$He$(e,e^\prime p)$$d$ cross section at
$\phi$=180 deg are compared to results of Glauber calculations using the
full $N$$N$ scattering amplitude (curve labeled \lq\lq full $F_{\rm lab}$\rq\rq), or its
central term only with a $\beta^{pp}$ parameter consistent with either
${\rm d}\sigma_{\rm el}^{pp}/{\rm d}t$ at small $t$ (curve labeled \lq\lq central
$F^{pp}$: small $\beta^{pp}$\rq\rq) or phase-shift analyses (curve labeled \lq\lq central
$F^{pp}$: large $\beta^{pp}$\rq\rq). All curves include two-body current contributions.
For the \lq\lq central $F^{pp}$: small $\beta^{pp}$\rq\rq calculation also shown are the results
obtained with one-body currents.}
\label{fig:xrycke1}
\end{figure}

\begin{figure}[bthp]
\includegraphics[width=6in]{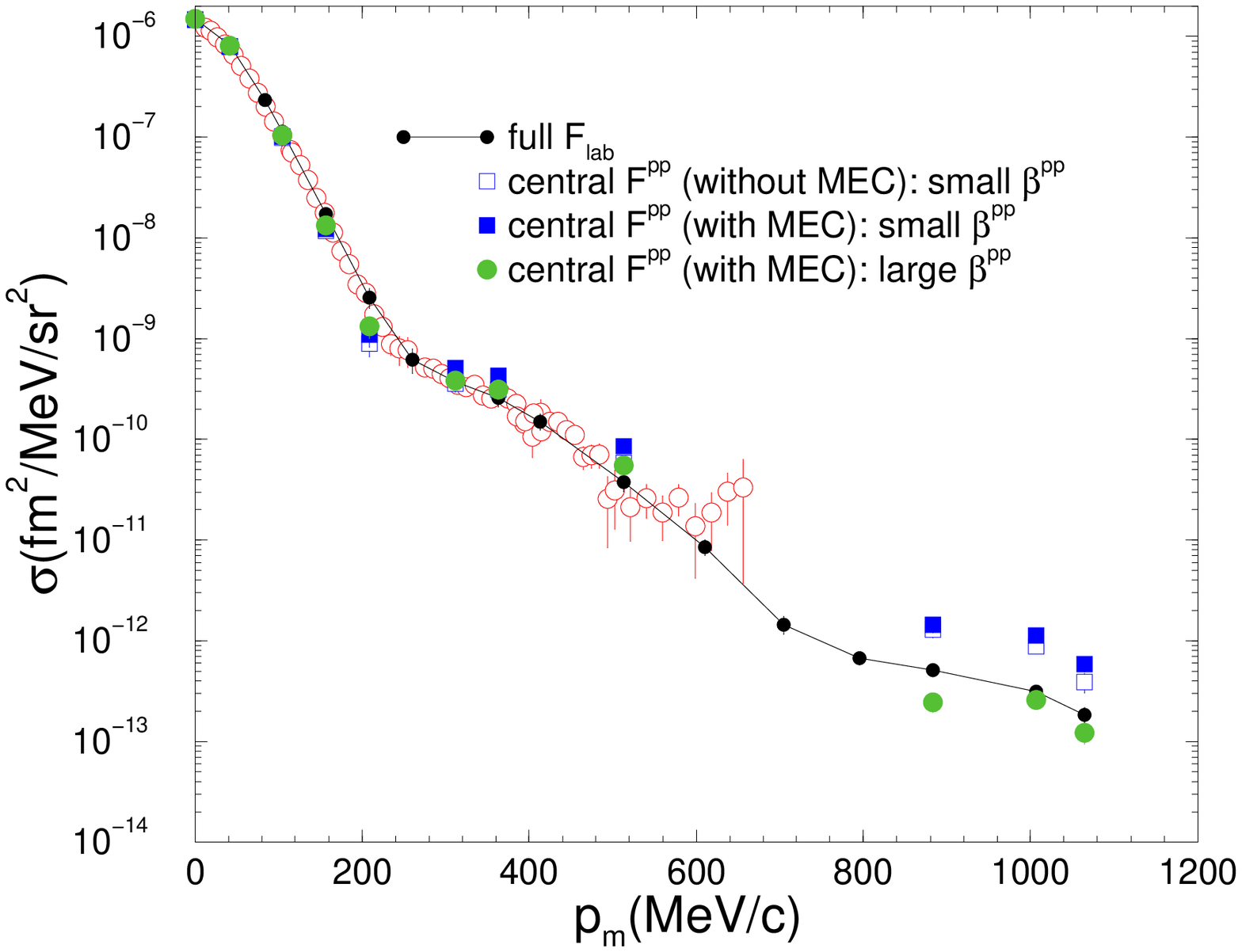}
\caption{Same as in Fig.~\protect\ref{fig:xrycke1}, but at $\phi$=0 deg.}
\label{fig:xrycke2}
\end{figure}

The sensitivity of the $^3$He($e,e^\prime p$)$d$ cross section, calculated
using these two different values for the $\beta^{pp}$ parameter (the isospin
dependence is ignored in the following discussion), is illustrated in 
Figs.~\ref{fig:xrycke1} and~\ref{fig:xrycke2}.  The curves labeled
\lq\lq central F$^{pp}$: small $\beta^{pp}$\rq\rq was obtained with $\beta^{pp}$= 
0.095 fm$^2$, as reported in Ref.~\cite{Ryckebusch03} (for $pp$ scattering).
The curve labeled \lq\lq central F$^{pp}$: large $\beta^{pp}$\rq\rq was obtained,
instead, with $\beta^{pp}$=0.157 fm$^2$, a value in line with that inferred from
phase-shift analyses~\cite{Wallace81}.  Incidentally, this large $\beta^{pp}$
was used in a set of unpublished calculations carried out by the present authors
in 2004 (and referred to in Refs.~\cite{Rvachev05,Ciofi05a,Laget05}).

The results of the calculation with a small $\beta^{pp}$, including only
one-body currents, are in quantitative agreement with those of Ref.~\cite{Ciofi05a},
although at $p_{\rm m} \simeq 400-700$ MeV/c the present predictions are
somewhat larger than obtained in Ref.~\cite{Ciofi05a}.  These relatively small
differences are presumably due to
i) the breakdown of the factorization approximation employed in
Ref.~\cite{Ciofi05a}, and ii) the use of the CC1 parameterization~\cite{deForest83}
adopted in Ref.~\cite{Ciofi05a}, rather than free nucleon form factors as in
the present study.

In the high $p_{\rm m}$ ($\simeq 1$ GeV/c) the small $\beta^{NN}$ results
lead to a large increase of FSI contributions.  The profile function
corresponding to a central scattering amplitude reads 

\begin{equation}
\Gamma^{NN}({\bf b};s)=
\frac{1}{8\pi\, \beta^{NN}(s)} \sigma^{NN}(s)\left[ 1-{\rm i} \rho^{NN}(s)\right]\,
{\rm exp}\left[-\frac{ {\bf b}^2}{4\beta^{NN}(s)} \right] \ ,
\end{equation}
and a small $\beta^{NN}$, while reducing its range, makes the value at zero
impact parameter larger, which leads to the FSI enhancement mentioned
above. 

Lastly, it is interesting to note that the results of the calculation
based on the full scattering amplitude derived from phase-shift analyses
are close to those using the central term only in the amplitude with
a $\beta^{NN}$ parameter obtained from the small $t$ slope of the elastic cross
section.
\section{Conclusions}
\label{sec:concl}

In this work we carried out calculations of the
$^3$He$(e,e^\prime p)$$d$ cross section for the kinematics
of JLab experiment E89-044, spanning the missing momentum range  
(0--1.1) GeV/c.  Final state interactions were treated in
the Glauber approximation, including both single and double
rescattering terms.  In contrast to earlier studies of the
same process~\cite{Laget05,Ciofi05a}, the profile operator
retained the full spin and isospin dependence of the underlying
$N$$N$ scattering amplitudes.  Parameterizations for these
were derived from phase-shift analyses in Ref.~\cite{Wallace81}.
It would be desirable to update and improve these parameterizations,
although the paucity of additional $N$$N$ scattering data
at energies beyond 500 MeV collected in the last two decades or so
(due in part to the termination of $N$$N$ programs at facilities
such as Saturne and LAMPF) will presumably not alter them significantly.

Corrections arising from boosting the amplitude from the c.m.~to the
relevant frame for the rescattering processes were
found to be relatively small.  However, the boosting procedure was
only implemented approximately.

Theory and experiment are in quantitative agreement for missing momenta
in the range (0--700) MeV/c.  Rescattering effects play a crucial role
over the whole range of $p_{\rm m}$, in particular double rescattering
processes are responsible for the increase of the cross section at $p_{\rm m}
\simeq 1$ GeV/c, which, nevertheless, is still underpredicted by theory by
about a factor of two.  However, the flattening of the data in this $p_{\rm m}$
region is well reproduced.  Two-body current contributions are relatively small,
but helpful in bringing theoretical predictions for the $A_{LT}$ observable
in significantly better agreement with experiment.  Spin-dependent terms in the
scattering amplitude are important at high $p_{\rm m}$.

A generalized Glauber approach has been recently developed for $(e,e^\prime p)$
processes~\cite{Frankfurt97,Sargsian05} which attempts to partially remove the
{\it frozen approximation} implicit in the original
derivation~\cite{Glauber59}.  The resulting correction
leads, in essence, to a modification of the profile operator
by a phase factor $\Gamma_{ij}\longrightarrow
{\rm exp}({\rm i} \Delta_0 z_{ij})\, \Gamma_{ij}$, with $\Delta_0$ fixed by
the kinematics, $\Delta_0=\omega\, E_{\rm m}/q$.  It was found to
be numerically very small, at the most 10\% at $p_{\rm m} \simeq 1$
Gev/c.  It may play a more prominent role, however, in the three-body
electrodisintegration $^3$He$(e,e^\prime p)$$p$$n$ at high missing energies.

Glauber calculations using only the central part of the scattering
amplitude with a $\beta^{NN}$ obtained by fitting the low $t$ slope of
${\rm d}\sigma_{\rm el}^{NN}/{\rm d}t$ are close to those using the
full amplitude derived from phase-shift analyses.  As argued in the
previous section, however, it is not clear that one is justified in
ignoring the spin dependence of the amplitude in view of the
large cross section differences observed in $N$$N$ scattering
involving polarized beam and target (see, for example,
Ref.~\cite{Lechanoine93}).
 
Future work aims at: i) extending the present Glauber approximation,
based on a spin- and isospin-dependent scattering amplitude,
to treat the $p$+$^3$H electrodisintegration of $^4$He and, in
particular, the reaction $^4$He$(\vec{e},e^\prime {\vec p}\,)$$^3$H, both
of which have been recently measured at JLab~\cite{Strauch03,Reitz05}
(the polarization parameters measured in the
$^4$He$(\vec{e},e^\prime {\vec p}\,)$$^3$H reaction~\cite{Strauch03}
have already been found to be in agreement with the results of a calculation
using an optical potential~\cite{Schiavilla05a}); ii) investigating 
the model dependence of the present predictions for the
$^3$He$(e,e^\prime p)$$d$ cross section upon the input (non-relativistic)
Hamiltonian adopted to generate the bound-state wave functions; iii) exploring
the extent to which the use of a relativistic Hamiltonian~\cite{Carlson93}
to generate these wave functions alters the present predictions, particularly
at high $p_{\rm m}$ (the research projects in items ii) and iii)
will be made possible by very recent developments of the hyperspherical-harmonics
method in momentum space~\cite{Kievsky05}); and iv) applying the
methods developed here to the three-body electrodisintegration of $^3$He.
Studies along these lines are being vigorously pursued.

\section*{Acknowledgments}

We would like to thank C.\ Ciofi degli Atti for illuminating correspondence
in regard to details of his and L.P.\ Kaptari's Glauber calculations,
J.\ Ryckebusch for providing parameterizations of $p$$p$ and $p$$n$
scattering amplitudes, and S.\ Jeschonnek, J.-M.\ Laget, and
M.M.\ Sargsian for interesting conversations.  We also would like to
thank V.R.\ Pandharipande and I.\ Sick for a critical reading of the
manuscript.

The work of R.S. was supported by DOE contract DE-AC05-84ER40150 under
which the Southeastern Universities Research Association (SURA) operates
the Thomas Jefferson National Accelerator Facility.  The calculations
were made possible by grants of computing time from the National Energy Research
Supercomputer Center.
\appendix
\section{}
\label{app:rp}

We list here explicit expressions for the single-proton response
functions $r^p_\alpha$ defined in Eq.~(\ref{eq:rpwia}):

\begin{eqnarray}
r_L^p &=&\left( \frac{q}{Q} \right)^2 \left[ G_{Ep}^2 + \frac{\eta}{1+\eta}
\left( \frac{p_\perp}{m} \right)^2
\left( G_{Mp}-\frac{1}{2} G_{Ep} \right)^2 \right] \ , \nonumber \\
r_T^p &=& \left( \frac{Q}{q} \right)^2 \left[ \frac{q^2}{2m^2} G_{Mp}^2
+\left( \frac{p_\perp}{m} \right)^2 \left[ \left( G_{Ep}+\frac{\eta}{2} G_{Mp} \right)^2
+\left( \frac{\omega}{4m} \right)^2 G_{Mp}^2 \right] \right] \ , \nonumber \\
r_{LT}^p &=& 2\sqrt{2} \, \left( \frac{p_\perp}{m} \right)
\left[ G_{Ep} \left( G_{Ep}+\frac{\eta}{2} G_{Mp} \right)
+\frac{q}{Q} \frac{\eta}{\sqrt{1+\eta}}
G_{Mp} \left( G_{Mp}-\frac{1}{2} G_{Ep} \right) \right] \ , \nonumber \\
r_{TT}^p &=& -\left( \frac{Q}{q} \right)^2 \left( \frac{p_\perp}{m} \right)^2
\left[ \left( G_{Ep}+\frac{\eta}{2} G_{Mp} \right)^2 - \left( \frac{\omega}{4m} \right)^2
G_{Mp}^2\right] \ , \nonumber
\end{eqnarray}
where $p_\perp$ is (the magnitude of) the component of the proton momentum
transverse to the momentum transfer ${\bf q}$, and $\eta$=$Q^2/(4m^2)$.  The
dependence of the proton electric ($G_{Ep}$) and magnetic ($G_{Mp}$) form
factors on $Q^2$ is understood. 
%
%
 
%
%
%
%
\clearpage
\clearpage
\end{document}